\documentclass{taira}

\usepackage{amssymb}
\usepackage{amsmath}
\usepackage{graphicx}
\usepackage{lipsum}
\usepackage{tabu}
\usepackage{multirow}

\begin{document}

\shorttitle{Unsteady control of supersonic cavity flow with resolvent analysis} 
\shortauthor{Q. Liu et al.} 

\title{Unsteady control of supersonic turbulent cavity flow based on resolvent analysis} 

\author{Qiong Liu\aff{1}
\corresp{\email{qliu3@ucla.edu}},
Yiyang Sun\aff{2}, 
Chi-An Yeh\aff{1}, 
Lawrence S.~Ukeiley\aff{3}, \\
Louis N.~Cattafesta III\aff{4}, 
\and
Kunihiko Taira\aff{1}
}

\affiliation{\aff{1}
Department of Mechanical and Aerospace Engineering, University of California, Los Angeles
CA 90095, USA 
\aff{2}
Department of Aerospace Engineering and Mechanics, University of Minnesota, Minneapolis, MN 55455, USA
\aff{3}
Department of Mechanical and Aerospace Engineering, University of Florida, 
Gainesville, FL 32611, USA
\aff{4}
Department of Mechanical Engineering, Florida State University, Tallahassee, FL 32310, USA
}

\maketitle

\begin{abstract}

We use resolvent analysis to determine an effective unsteady active control setup to attenuate pressure fluctuations in turbulent supersonic flow over a rectangular cavity with a length-to-depth ratio of $6$ at a Mach number of $1.4$ and a Reynolds number based on cavity depth of $10,000$.  
Large-eddy simulations (LES) and dynamic modal decomposition (DMD) of the supersonic cavity flow reveal the dominance of two-dimensional Rossiter modes II and IV.  
These predominantly two-dimensional vortical structures generate high-amplitude unsteadiness over the cavity through trailing-edge impingement and create oblique shock waves by obstructing the freestream.   
To disrupt the undesired formation of two-dimensional spanwise vortical structures, we introduce three-dimensional unsteady forcing along the cavity leading edge to generate streamwise vortical structures.  
Resolvent analysis with respect to the time-averaged base flow is leveraged to determine the optimal combination of forcing frequency and spanwise wavenumber.  
Instead of selecting the most amplified resolvent forcing modes, we seek the combination of control parameters that yields sustained amplification of the primary resolvent-based kinetic energy distribution over the entire length of the cavity.  
The sustained amplification is critical to ensure that the selected forcing input remains effective to prevent the formation of the large spanwise vortices.  
This resolvent-analysis-based flow control guideline is validated with a number of companion LES of the controlled cavity flows. 
As guided by the kinetic energy metric, the optimal control setup is verified to be the most effective in reducing the pressure root-mean-square level up to $52\%$ along the aft and bottom cavity walls compared to the baseline cavity flow.  
The level of reduction in pressure fluctuation is approximately double of what can be achieved by comparable steady blowing based flow control.   
We also perform dynamic mode decomposition on the controlled flows and find that the actuation input indeed suppresses the formation of two-dimensional large-scale Rossiter modes.  
The present flow control guideline derived from resolvent analysis should be applicable to flows that require the actuation input to remain effective over an extended region of interest.

\end{abstract}

\section{Introduction}
\label{sec:intro}

Flows over open cavities are comprised of rich physics owing to the natural feedback loop. Disturbances arise from the cavity leading edge and form large vortical structures in the shear-layer region due to the amplification of Kelvin--Helmholtz instabilities.  
These large-scale structures impinge on the cavity trailing edge, giving rise to strong acoustic waves and intense pressure fluctuations. The acoustic waves propagate upstream within the cavity, instigate the Kelvin--Helmholtz instabilities in the shear layer and produce a self-sustained feedback oscillation.
In the case of supersonic cavity flow, the fluctuations are accentuated by the appearance of shock waves, which can cause structural damage \citep{mcgregor1970drag}. 
Because cavity flows appear in a wide range of engineering applications, including landing gear wells, aircraft bays, automotive sun roofs, and high-speed trains, there is a strong interest in attenuating the pressure fluctuations over the cavity with flow control techniques \citep{cattafesta2008active, rowley2006dynamics}.

A large number of studies have focused on analyzing the shear layer within cavity flow in an attempt to explain the cavity flow unsteadiness. \cite{krishnamurty1955acoustic} described the resonant oscillations associated with acoustic tones in cavity flows.  Shortly thereafter, \cite{Rossiter1964} derived the well-known semi-empirical formula to predict the frequencies of the anharmonic resonant tones in cavity flows, which was slightly modified by \cite{heller1971flow}. Extensive research has since been carried out to understand the amplitude and other characteristics of cavity flows for different cavity geometries, incoming flow conditions, and Reynolds numbers \citep{rockwell1979self, colonius2001overview, rowley2006dynamics, lawson2011review}.

As visible flow structures emerge in cavity flows, modal analysis techniques have been applied to extract the dominant flow features and to understand the underlying flow mechanisms \citep{Taira:AIAAJ17, Taira:AIAAJ20}. 
Subsonic cavity flow fields obtained from experiments have been examined by \cite{murray2009properties} with a purely spatial application of the Proper Orthogonal Decomposition. Their analysis revealed that the most energetic modes are the shear-layer modes, whose spatial structures remain similar in the subsonic regime.  In addition to the primarily two-dimensional shear layer mode associated with the Kelvin--Helmholtz instability, a three-dimensional centrifugal instability stemming from the large recirculation inside the cavity has been analyzed with experimental \citep{plumblee1962theoretical,maull1963three, faure2007visualizations, Faure:EF2009, larcheveque2007large} and numerical approaches \citep{bres2008three, de2014three}. \cite{faure2007visualizations} revealed the evidence of three-dimensional structures inside of the cavity. They showed that the dynamics of the structures were not due to the secondary shear layer instabilities. \cite{bres2008three} used biglobal instability analysis \citep{theofilis2011global} to examine three-dimensional (spanwise) instability of the two-dimensional mean cavity flow.
The identified spanwise (centrifugal) instability within the cavity possesses a frequency that is an order of magnitude lower than those associated with the two-dimensional shear-layer instabilities. 
Recent studies have examined the influence of Mach number and sidewalls on the oscillations of the cavity \citep{Beresh:JFM2016, sun2017biglobal, liu2016linear, picella2018successive, casper2018spatial}.

Various passive and active control strategies have been developed with varying degrees of success to suppress the dominant oscillations \citep{cattafesta2008active}.  
A few specific examples out of the numerous studies are highlighted below to demonstrate the persistent need for physics-based studies to guide flow control actuation design.    
\cite{ukeiley2004suppression} performed an experiment with a small suspended cylinder at the leading edge of the cavity following the work of \cite{mcgrath1996active}. The results showed that lifting the shear layer away from the cavity led to significant suppression of the pressure fluctuations, but the performance was highly dependent on the placement of the cylinder. 
\cite{sarno1994suppression} used a static fence along the leading edge of the cavity to manipulate the shear layer and suppress the cavity flow oscillations. 
However, the magnitude of suppression was again highly dependent on the modal frequency and the flow conditions.  This phenomenon of suppression degradation associated with a deviation from the design condition is a common characteristic of passive flow control strategies.

Active flow control can provide adaptive capability over a wide range of flow conditions \citep{colonius2001overview,rowley2006dynamics, cattafesta2011actuators}. 
\cite{vakili1994control} experimentally studied the effectiveness of steady injection from the leading edge to control the unsteadiness in supersonic cavity flow. 
Moreover, recent studies have investigated the effect of spanwise spatial variations of steady blowing, which showed the three-dimensional steady injection surpasses its two-dimensional counterpart in terms of suppression and efficiency \citep{lusk2012leading, george2015control, zhang2019suppression, sun2018effects}.
\cite{rizzetta2003large}, for example, performed a number of LES on the suppression of cavity flow oscillations using pulsed injection at very high frequency. 
This type of control suppresses resonant acoustic oscillations with even lower energy input compared to steady injection, but high-frequency forcing in high-speed flows remains a challenge for actuators \citep{cattafesta2011actuators}.
While there have been a number of similar studies on unsteady control of cavity flows, there has not been a systematic investigation on the input-output characterization of high-speed cavity flows with a focus on forcing frequency and actuator placement.

To shed light on these open questions, the present study considers the use of resolvent analysis \citep{trefethen1993hydrodynamic, jovanovic2005componentwise} with respect to the time-mean flow \citep{mckeon2010critical}. \cite{nakashima2017assessment} capitalized on resolvent analysis to elucidate the control mechanism of the sub-optimal control on coherent structures in the wall-bounded turbulent flows. \cite{yeh2018resolvent} used the insights from resolvent analysis to design active thermal actuation to suppress the flow separation over an airfoil. 
They used the response mode and gain from the resolvent analysis to form a metric to quantify momentum mixing and to select control parameters. The chosen control parameters were verified to deliver effective separation control with companion LES.

Resolvent analysis has also been used by \cite{leclercq2019linear} to design a closed-loop strategy to suppress oscillations in two-dimensional laminar cavity flow.  They achieved effective suppression of oscillations over the cavity.  However, additional progress is required to develop resolvent analysis based flow control approaches for high-speed turbulent cavity flow, which possesses complex physics beyond the two-dimensional, laminar, incompressible flow. In the present study, we demonstrate the use of resolvent analysis as a physics-based approach to systematically design an effective open-loop control technique for supersonic turbulent cavity flows with rich three-dimensional, turbulent and shock dynamics. A design metric is introduced here as a predictive norm to identify effective flow control approaches.  Companion LES are performed with the selected control parameters for assessing the chosen flow control setups.  We also analyze the underlying control mechanisms by studying the coherent structures using dynamic mode decomposition \citep{schmid2010dynamic, rowley2006dynamics}.

The current paper is organized as follows. We present in \S\ref{sec:problem} the details of the open cavity flow problem, the open-loop control setup, and the resolvent analysis framework.  Turbulent supersonic cavity flow characteristics, the properties of the response and forcing modes from resolvent analysis, and three-dimensional dynamic coherent structures are presented and discussed in \S\ref{sec:baseline} to identify effective control parameters for unsteady cavity open-loop flow control.  Based on the insights gained from resolvent analysis, we then perform a series of controlled cavity flow simulations,
illustrate the underlying control mechanism, 
and correlate the control effect to the resolvent mode in \S\ref{sec:control}.  
Finally, concluding remarks are offered in \S\ref{sec:summary}.


\section{Approaches}
\label{sec:problem}

\subsection{Problem setup}

We consider spanwise-periodic supersonic turbulent flows over a rectangular cavity with a length-to-depth ratio of $L/D = 6$ at a free-stream Mach number of $M_\infty \equiv u_\infty/a_\infty=1.4$, where $a_\infty$ is the sound speed and $u_\infty$ is the free-stream velocity.  The Reynolds number based on the cavity depth is $\Rey \equiv \rho_\infty u_\infty D/\mu_\infty=10,000$ with $\mu_\infty$ and $\rho_\infty$ being the free-stream dynamic viscosity and density, respectively.  In the present study, all variables are nondimensionalized; namely, lengths by the cavity depth $D$, temperature by $T_\infty$, pressure by $\frac{1}{2} \rho_\infty u_\infty^2$, density by $\rho_\infty$, and time by $D/u_\infty$. The initial boundary layer thickness at the leading edge is set to be $\delta_0/D=0.167$.  The domain has spanwise periodic boundary conditions imposed with a spanwise periodicity of $W/D = 2$.  

We study the turbulent cavity flow with LES using a compressible flow solver \emph{CharLES} \citep{Khalighi:ASME2011,Bres:AIAAJ17}. The solver is based on a second-order finite-volume discretization and a third-order Runge--Kutta time integration scheme.  For the present LES, Vremen's sub-grid scale model \citep{Vreman:PF04} is utilized and the Harten-Lax-van Leer contact scheme \citep{Toro:94} is used to capture shocks in the supersonic flows.  Random Fourier modes are superimposed to the one seventh power law velocity profile to simulate unsteady fluctuations from the inlet \citep{bechara1994stochastic}. No-slip and adiabatic wall boundary condition is specified along the cavity walls. Sponge boundary condition is applied in the farfield and outlet boundaries for damping out exiting waves and preventing numerical reflections.

\begin{figure}
\centering
\includegraphics [width=0.8\textwidth,trim={0 0cm 0 0cm},clip]{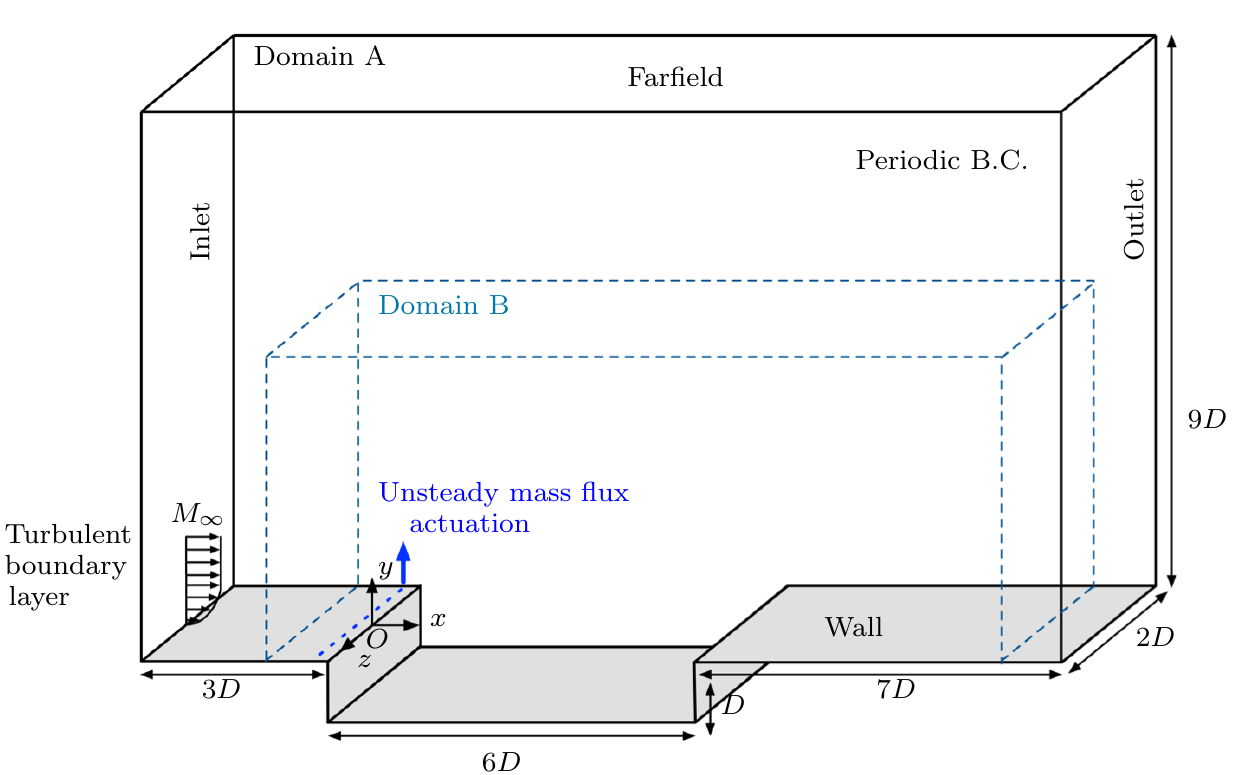}
\caption{Setup for computational domain and boundary conditions. Domain A is used for the baseline and controlled LES cases.  Domain B is used for resolvent and dynamic mode decomposition analyses.}
\label{fig: geo}	
\end{figure}

The computational setup for the present study is shown in figure \ref{fig: geo}. A Cartesian coordinate system is used with its origin placed at the spanwise center of cavity leading edge, with $x$, $y$ and $z$ denoting the streamwise, wall-normal and spanwise directions, respectively.  The computational domain extends upstream and downstream by 3 and 7 times of the cavity depth, respectively. The far-field extent is set to be 9 times of the cavity depth. The computational domain is discretized with a structured mesh with 14 million cells for the baseline and 16 million cells for the controlled simulations, for which mesh refinement is applied in the vicinity of the actuators.  This computational setup has been verified to be sufficient to characterize flow properties as reported in our previous work \citep{sun2018effects}.
A small three-dimensional domain, labeled as Domain B in figure \ref{fig: geo}, is used for the modal analyses. It has a upstream length equal to the cavity depth. The downstream and far-field boundaries are placed 5 times of cavity depth away. The grid size for Domain B is approximately 70,000 cells.

\subsection{Unsteady actuation setup}

We perform active control of the cavity flow by introducing blowing and suction along the leading edge of the cavity through a mass flux boundary condition, as illustrated by the blue dashed line in figure \ref{fig: geo}. The actuator is located at 
$x_c =-0.0698$ with a streamwise slot extent of $\Lambda_c =0.0175$ following our previous experimental studies \citep{lusk2012leading, george2015control, zhang2019suppression}. The unsteady actuation is prescribed with a wall-normal velocity profile of 
\begin{equation}\label{eqn:actuation}
    u_\text{jet}(x,z,t)
    = A\sin(\omega_c t)
    \Phi(x,x_c,\Lambda_c)\cos(\beta_c z),
\end{equation}   
where $A$ is the actuation amplitude, $\omega_c$ and $\beta_c$ are the actuation frequency and spanwise wavenumber, respectively.  In what follows, we report the actuation frequency $\omega_c$ through its dimensionless cavity length-based Strouhal number $St_c=\omega_c L/(2\pi u_\infty)$. 
The spatial velocity profile for the actuator is given by
\begin{equation}
    \Phi(x,x_c,\Lambda_c)
    = \frac{1}{4}\left\{1+\tanh\left[\kappa_1(x-x_c+\Lambda_c/\kappa_2)\right]\right\} 
    \left\{1-\tanh\left[\kappa_1(x-x_c-\Lambda_c/\kappa_2)\right]\right\}
\end{equation} 
along the streamwise direction to avoid the velocity discontinuity at the edge of the actuator.  Here, we choose $\kappa_1 = 2000$ and $\kappa_2 = 2.6$. The actuation frequency and spanwise wavenumber will be selected based on the resolvent analysis as discussed later in \S\ref{sec:Resolvent_methodology}.

The actuation efforts in this study are reported in terms of the unsteady momentum coefficient defined by
\begin{equation}
    C_\mu' \equiv \frac{J}{\frac{1}{2} \rho_\infty u^2_\infty W\delta_0}, 
\end{equation}
where $J = (\rho_\infty \omega_c/2\pi) \int_{T_c} \int_S u_\text{jet}(x,z,t)^2 {\rm d}s {\rm d}t$, $T_c=2\pi/\omega_c$ is the period of the unsteady control actuation, and $S$ is the actuation area. Here, we normalize the unsteady control effect with the boundary layer thickness $\delta_0$ and set the unsteady momentum coefficient to be $C_\mu'=0.02$, following canonical values from past unsteady control studies \citep{elimelech2011secondary,shaw1998active,williams2007supersonic}.

The control effects are assessed using the surface integrated pressure root-mean-square on the aft and bottom walls ($\tilde{\Omega}$) defined as
\begin{equation}\label{eqn:assess}
\Delta \tilde{p}_\text{rms}=\frac{(\tilde{p}_\text{rms,c}-\tilde{p}_\text{rms})}{\tilde{p}_\text{rms}}, 
\quad \text{where} \quad 
\tilde{p}_\text{rms} = \int_{\tilde{\Omega}} \frac{p_\text{rms}}{\frac{1}{2}\rho_\infty u^2_\infty}\text{d}S
\end{equation}
and $\tilde{p}_\text{rms,c}$ is the surface integrated pressure fluctuation in the controlled cavity flows. The global $\Delta \tilde{p}_\text{rms}$ quantifies the level of pressure fluctuations over the cavity.

\subsection{Resolvent analysis}\label{sec:Resolvent_methodology}

Resolvent analysis is used to reveal the flow response to harmonic forcing input with respect to a given base state.  In the analysis, we assume the flow $ q(\boldsymbol{x},t)$ can be expressed through Reynolds decomposition 
\begin{equation}\label{eqn:decompostion}
q(\boldsymbol{x},t)={\bar q}(\boldsymbol{x})+  q'(\boldsymbol{x},t),
\end{equation}
where ${\bar q}(\boldsymbol{x})$ is the time-invariant base state and $q'(\boldsymbol{x},t)$ is the fluctuation.  Traditionally, the base state ${\bar q}$ is taken to be an equilibrium state \citep{trefethen1993hydrodynamic,trefethen2005spectra,jovanovic2005componentwise}.  In the present study, we consider the time-averaged turbulent cavity flow as the base state with fluctuations being statistically stationary \citep{mckeon2010critical}.

By substituting the Reynolds decomposed state variable (\ref{eqn:decompostion}) into the Navier--Stokes (NS) equations, the governing equation for the fluctuation $q'$ becomes
\begin{equation}\label{eqn:decom_NS}
    \frac{\partial q'}{\partial t}=L(\bar q)q'+f,    
\end{equation}
where $L(\bar q)$ denotes the linearized NS operator about $\bar{q}$. Here, $f$ is viewed as the forcing input comprised of the sum of remaining terms including the nonlinear terms, which can be interpreted as internal forcing in turbulent flow about the base state due to nonlinear interactions \citep{mckeon2010critical}.

In the present study, we consider the three-dimensional turbulent cavity flow with spanwise periodic boundary conditions.  Hence, the time- and spanwise-averaged flow is used as the base state $\bar q$. The fluctuation $q'$ and forcing $f$ are expressed as Fourier modes with spanwise wavenumber $\beta$ and frequency $\omega$, such that
\begin{equation}
q'(\boldsymbol{x},t) =\hat{q}(x, y) e^{\text{i}(\beta z + \omega t)},
\quad 
f(\boldsymbol{x},t) = \hat{f}(x,y)e^{\text{i}(\beta z+\omega t)},
\label{eqn:model_expression}
\end{equation}
where frequency $\omega$ and spanwise wavenumber $\beta$ are real numbers.

Substituting the modal expressions (\ref{eqn:model_expression}) into (\ref{eqn:decom_NS}), we can find the relationship between the forcing $\hat{f}(x,y)$ and the fluctuation $\hat{q}(x,y)$ for each combination of $(\omega, \beta)$:
\begin{equation}\label{eqn: Resolvent}
	\hat q_{\omega,\beta} 
	= [\text{i}\omega I - L(\bar q, \beta)]^{-1}\hat f_{\omega,\beta}
	= H(\bar q; \omega, \beta)\hat f_{\omega,\beta}. 
\end{equation}
Here, the operator $H(\bar q; \omega, \beta) \equiv [\text{i}\omega I - L(\bar q, \beta)]^{-1}$ is referred to as the {\it resolvent} operator, which serves as the transfer function between the forcing (input) $\hat f_{\omega,\beta}$ and the response (output) $\hat q_{\omega,\beta}$ about the base state $\bar{q}$ for the given frequency $\omega$ and spanwise wavenumbers $\beta$ \citep{jovanovic2005componentwise, mckeon2010critical, schmid2012stability}.


Since the resolvent operator serves as the transfer function between the output response and the input forcing, the stability property of the transfer function, more specifically the linear operator $L(\bar{q},\beta)$, must be evaluated first.  
We perform stability analysis on the linear operator $L(\bar{q},\beta)$ to examine if the flow system is stable or unstable.  The analysis forms an eigenvalue problem as
\begin{equation}
L(\bar{q};\beta)\hat{q}_\beta=\lambda \hat{q}_\beta ,
\end{equation}
where $\lambda=\lambda_r+\text{i} \lambda_i $ with $\lambda_i$ representing the temporal frequency of the perturbation and $\lambda_r$ being the growth ($\lambda_r>0$) or decay ($\lambda_r<0$) rate of the perturbation.

For a stable base state ($\lambda_r<0$ for all $\lambda$), 
we can directly perform resolvent analysis to examine energy amplification of the flow system to the harmonic forcing over an infinite-time horizon. However, for an unstable base state, care is needed to obtain meaningful physical insights from resolvent analysis.  
In this case, a perturbation will grow exponentially from the base state due to the instability (although the fluctuations will saturate in the full nonlinear flow).  As such, exponential growth can be larger than the resolvent-based amplification. We provide a remedy to this issue by performing resolvent analysis over a finite-time window.  To accomplish this, we use exponential discounting \citep{jovanovic2004modeling,yeh2018resolvent} for handling the unstable base flows, such as the time-averaged turbulent cavity flow in the present study. Incorporating discounting into the resolvent operator can be considered as constraining the system response within a short period of time before it diverges from the unstable base state.

We perform the discounted resolvent analysis for the unstable base state by applying an exponential temporal filter $\exp{(-\kappa t)}$ with $\kappa \ge 0$ to equation (\ref{eqn:decom_NS}), which yields  \begin{equation}
    \mathcal{H}(\bar{q};\kappa, \omega, \beta)
    =[(\kappa + \text{i}\omega) I- L(\bar q; \beta)]^{-1}.
\end{equation}
The magnitude of $1/\kappa$ characterizes this finite-time window. For an unstable system with maximum growth rate $\max(\lambda_r)> 0$, the discounting parameter is chosen such that $\kappa > \max(\lambda_r)$ to impose an appropriate finite-time window for the unstable system.

To assess the optimal energy amplification of the system, the resolvent analysis is cast in the framework of singular value decomposition (SVD) of the resolvent operator $H$ under the compressible energy norm \citep{chu1965energy} 
\begin{equation}
	E=\int_S \left[ 
	\frac{\bar{a}^2 \rho^2}{\gamma \bar{\rho}}+\bar{\rho}({u}^2+{v}^2+{w}^2) + \frac{\bar{\rho}C_v T^2}{\bar{T}}
	\right] {\text d}s,
\end{equation}
which yields
\begin{equation}\label{eqn: Resolvent_2}
	W^{\frac{1}{2}}H(\bar q; \omega, \beta)W^{-\frac{1}{2}} = Q \Sigma  F^\ast,
\end{equation}
where $W$ is the weight matrix based on the compressible energy weight above and $S$ is the area of analysis as illustrated in figure \ref{fig: geo}.  
The matrix $Q=[\hat{q}_1, \hat{q}_2,\dots, \hat{q}_n]$ holds the set of response modes and the right matrix $F=[\hat{f}_{1}, \hat{f}_{2},\dots, \hat{f}_{n}]$ contains the set of forcing modes. 
The superscript $\ast$ denotes the Hermitian transpose. 
The singular values $\Sigma=\text{diag}(\sigma_1,\sigma_2,\dots,\sigma_n)$ represent the amplification between the response and forcing modes in descending order.  Additional details on the framework of the present resolvent analysis and cavity flows can be found in our companion work \citep{sun2019resolvent}.


\section{Uncontrolled cavity flow}
\label{sec:baseline}

\subsection{Baseline flow characteristics}

\begin{figure}
    \centering
    \includegraphics [width=0.9\textwidth,trim={0 0cm 0 0cm},clip]{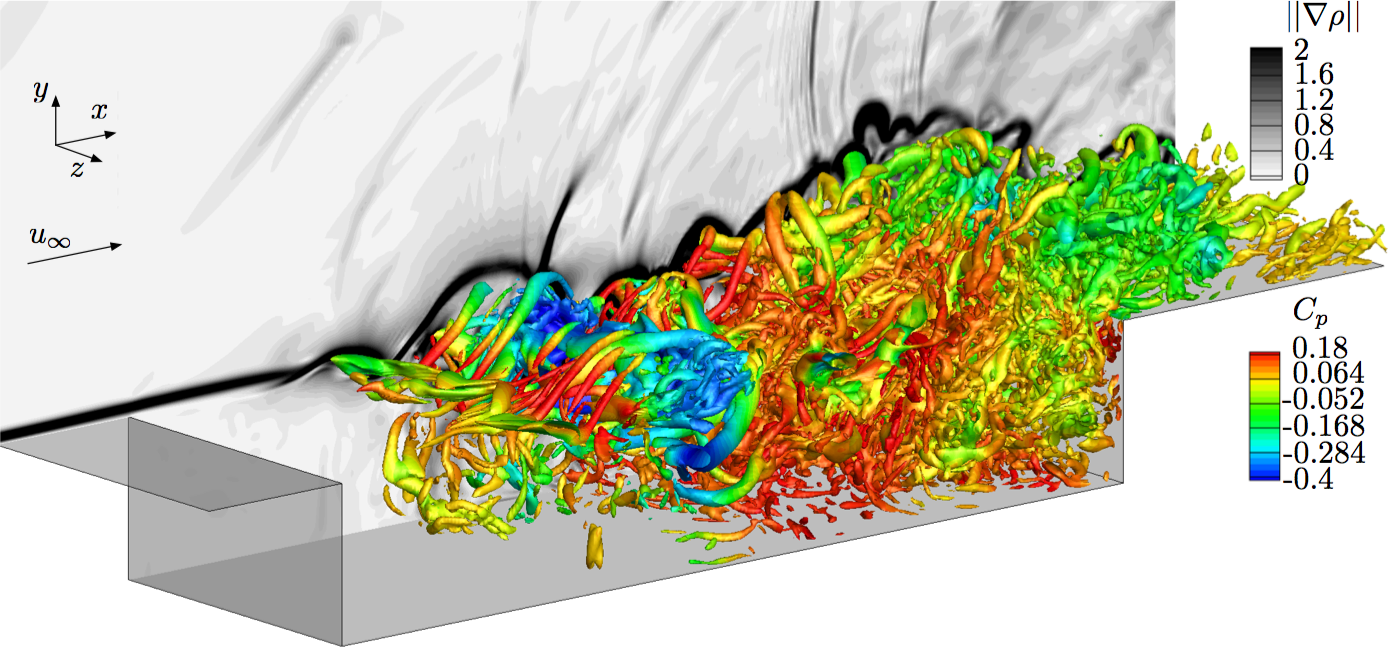}
    \caption{Instantaneous cavity flow visualized with the $Q$-criterion $Q(D/u_\infty)^2$=10 colored by $C_p=(p-p_\infty)/(\frac{1}{2}\rho_\infty u^2_\infty)$ at $M_\infty=1.4$. Numerical schlieren is shown in the background in gray scale.}
    \label{fig: baseline_vortices}
\end{figure}

The baseline supersonic turbulent flow over the rectangular cavity at $M_\infty=1.4$ and $\Rey=10,000$ will be described first.  Figure \ref{fig: baseline_vortices} shows the instantaneous flow with its vortical structures visualized using the isosurface of $Q$-criterion colored by the pressure coefficient.  The background shows the numerical schlieren with $\| \nabla \rho \|$ to capture the compression waves.  The incoming flow forms a shear layer that emanates from the leading edge of the cavity. The shear layer rolls up into large spanwise vortices at around a third of the cavity length, generating large fluctuations over the cavity.  Once the large spanwise vortices reach the middle of the cavity, there is an observable loss of spanwise coherence and the emergence of small-scale vortical structures. In supersonic cavity flows, the large-scale structures obstruct the free stream and create compression waves.  Moreover, the impingement of these vortical structures onto the aft cavity wall generates strong waves that travel upstream within the subsonic region.  The vortical and pressure fluctuations produce a high level of unsteadiness in and above the cavity.  Towards the aft of the cavity, there is a complex interplay between the shear layer, the impingement of the vortical structures on the aft wall, and the recirculation within the cavity.

\begin{figure}
\centering
\includegraphics [width=0.49\textwidth,trim={0 0cm 0 0cm},clip]{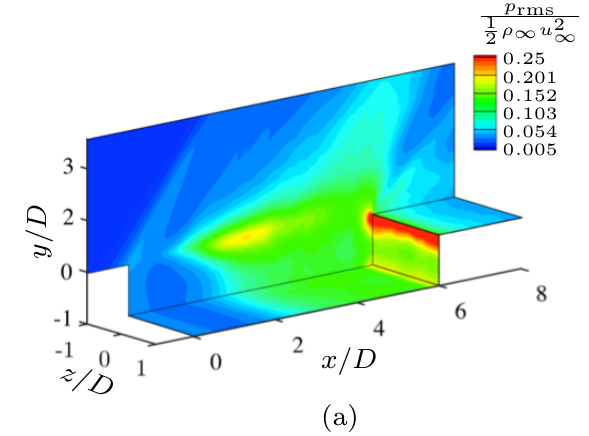}
\includegraphics [width=0.49\textwidth,trim={0 0cm 0 0cm},clip]{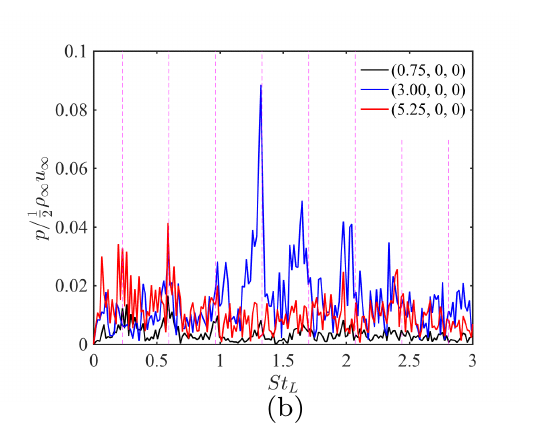}
\caption{(a) The rms pressure, $p_\text{rms}/(\frac{1}{2}\rho_\infty u^2_\infty)$, along the cavity surfaces and $x-y$ plane along $z/D=-1$. (b) Frequency spectra of instantaneous pressure $p/(\frac{1}{2}\rho_\infty u^2_\infty)$ at three probe locations along the shear layer.}	
\label{fig:b_spectral}
\end{figure}

High levels of unsteadiness emerge along the aft wall due to the impingement of vortical structures and within the shear layer from the formation of the large-scale vortical structures.  The pressure fluctuations in terms of its normalized root mean square (rms) are presented in figure \ref{fig:b_spectral}(a).  Here, the strong rms levels along the cavity trailing edge are seen, in the shear layer near the middle ($2\lesssim x/D \lesssim 4$) of the cavity and above the cavity from the unsteady compression waves.

Spectral properties of the fluctuations at three representative locations of $(x, y, z)/D =(0.75, 0, 0)$, $(3, 0, 0)$ and $(5.25, 0, 0)$ along the shear layer are discussed here.  The frequency spectra of the pressure measurements at these three probe locations are shown in figure \ref{fig:b_spectral}(b).  The dominant oscillation frequency shifts as the flow advects over the cavity.  We can compare the frequency spectra with the resonant peaks predicted by the modified Rossiter's semi-empirical formula \citep{heller1971flow}  
\begin{equation}\label{eqn:Rossiter}
 St_L=\frac{fL}{u_\infty}
 =\frac{n-\alpha}{1/k + M_\infty /\sqrt{1+(\gamma-1)M_\infty^2/2}},   
\end{equation}
where specific heat ratio of $\gamma=1.4$, the average convective speed of disturbance in shear layer $k=0.57$, and the phase delay $\alpha=0.25$ for the $n$th Rossiter mode.  At $x/D=0.75$, the primary oscillation appears at $St_L=0.6$ with a low amplitude.  Once the flow reaches $x/D=3$, the dominant frequency corresponds to Rossiter mode IV, whose Strouhal number is $St_L=1.33$.  At the trailing edge, the dominant frequency shifts to the lower Rossiter mode II frequency $St_L=0.6$, with a substantial reduction in the oscillation amplitude.  The shift in the dominant oscillation frequency as the vortical structures convect over the cavity suggests strong nonlinear interactions in the flow.

\begin{figure}
\centering
\includegraphics [width=\textwidth]{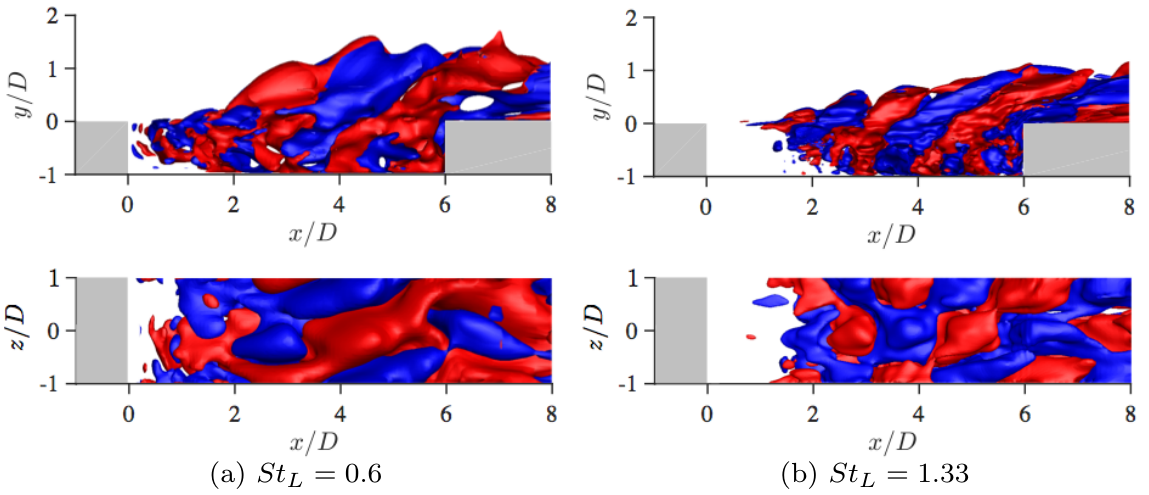}
\caption{Spatial structures of DMD modes visualized with real streamwise velocity iso-surface $(\pm 0.3)$.}	
\label{fig:dmd_mode}
\end{figure}

To identify the flow structures corresponding to the peak frequencies, we perform DMD \citep{schmid2010dynamic, rowley2009spectral} on the snapshots of the flow field to extract the three-dimensional coherent structures associated with the detected frequencies. 
Isosurfaces of the streamwise velocity component of the DMD modes at the frequencies of $St_L=0.6$ and $1.33$ are shown in figure \ref{fig:dmd_mode}.   The DMD modes capture the shear-layer modes associated with the respective single frequency.  By comparing the spatial structures for these two frequencies, it is observed that the DMD mode for $St_L=0.6$ possesses larger structures in both the streamwise and vertical directions.  In contrast, the structures in the DMD mode for $St_L = 1.33$ are approximately half in size in the streamwise direction.  Noteworthy here is that these two DMD (shear layer) modes exhibit spatial variations with the same the spanwise wavenumber of $\beta=\pi$ as evident from the top views in figure \ref{fig:dmd_mode}.

\subsection{Resolvent spectra and modes}
\label{sec:resolvent}

\begin{figure}
\centering
\includegraphics [width=\textwidth]{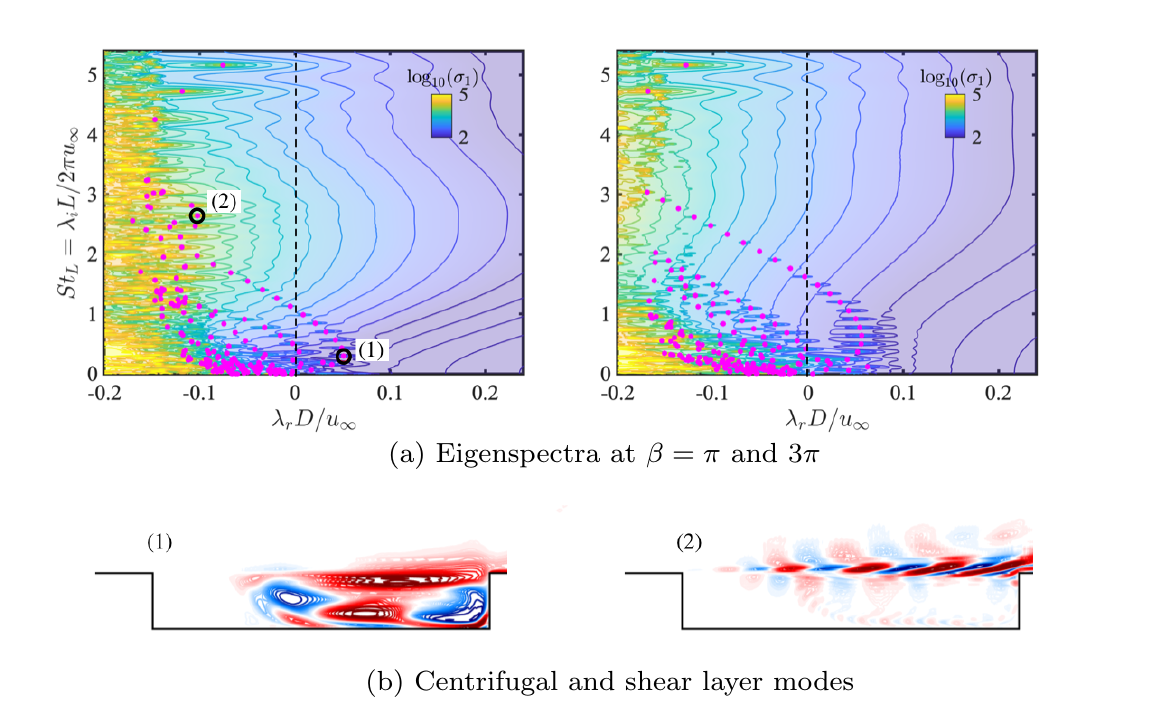}
\caption{(a) Eigenspectra and pseudospectra of $L(\bar{q};\beta)$ for spanwise wavenumbers $\beta=\pi$ (left) and $3\pi$ (right). The magenta dots indicate the growth rates and frequencies of $L(\bar{q};\beta)$. The dashed line indicates the neutral stability line. The contour plots visualize the pseudospectra. (b) The (1) centrifugal and (2) shear-layer modes visualized by the real part of the streamwise velocity. The locations of their eigenvalues are indicated on plot (a).}
\label{fig:global}
\end{figure}

Stability analysis is performed on $L(\bar{q}; \beta)$ to evaluate its stability properties. 
The eigenspectra are shown in figure \ref{fig:global} for two representative cases with $\beta=\pi$ and $3\pi$. 
The imaginary part of the eigenvalue $\lambda$ is normalized as $\lambda_i L/(2\pi u_\infty)$ and the real part is normalized as $\lambda_r D/u_\infty$. The results indicate the linear operator $L(\bar{q}; \beta)$ is unstable with positive growth rates ($\lambda_r D/u_\infty > 0$) of eigenmodes, including both centrifugal and shear-layer modes. 
The spatial structures of two representative unstable modes, centrifugal mode, and shear layer mode, are shown in figure \ref{fig:global} (b). The centrifugal mode is present inside of the cavity, which is relevant to the recirculation motion \citep{bres2008three}. However, the shear layer mode appears over the cavity and is formed due to the Kelvin-Helmholtz instability.

We show the pseudospectra contours \citep{trefethen2005spectra} of the linear operator $L(\bar q;\beta)$ as the background in figure \ref{fig:global}. Higher values of contours appear on the left-hand side planes for the cases of $\beta=\pi$ and $3\pi$, where the stable eigenvalues are located. For the case of $\beta=\pi$, pseudospectral levels protrude far into the unstable plane at $1<\lambda_i L/(2\pi u_\infty) < 4$ representing a non-normal behavior of the operator. While for the case of $\beta=3\pi$, the non-normal behavior of operator occurs in the frequency range of $2<\lambda_i L/(2\pi u_\infty) < 5$. The non-normal behavior of the operator can cause the flow to exhibit significant energy amplification.

\begin{figure}
\centering
\includegraphics [width=0.48\textwidth]{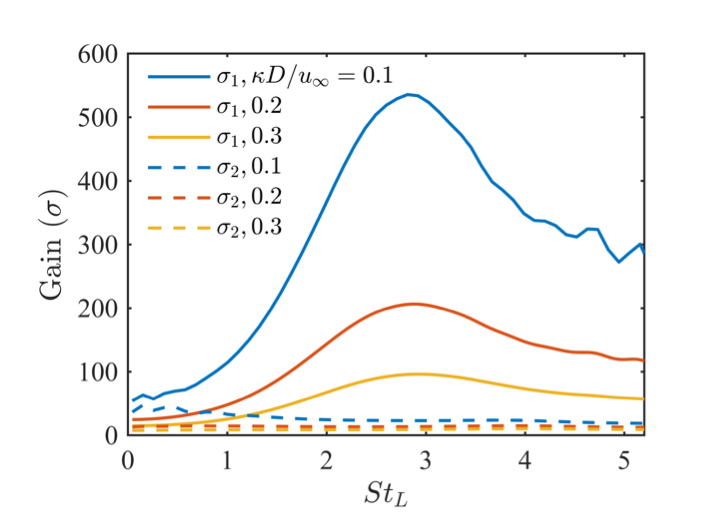}
\caption{
Resolvent gain calculated using discounting parameter of $\kappa D/u_\infty= 0.1$, $0.2$ and $0.3$ for $\beta=2\pi$. }	
\label{fig:resolvent_gain_kappa}
\end{figure}

Because the linear operator $L(\bar{q}; \beta)$ is unstable, we perform a discounted resolvent analysis.
The choice of the discounting parameter is associated with temporal windowing of the system response to examine the dynamics before the instability diverges away, as discussed in \ref{sec:Resolvent_methodology}. 
Figure \ref{fig:resolvent_gain_kappa} shows the leading and secondary gains for the case of $\beta=2\pi$ using different discounting parameters of $\kappa D/u_\infty=0.1$, $0.2$ and $0.3$.  We observe that as $\kappa D/u_\infty$ increases, the magnitude of resolvent gain decreases.  However, the profile of the gain distribution is not altered with the different choice of discounting parameter.  In fact, the maximum gain is achieved at the same Strouhal number of $St_L\approx3$.

The large separation between the leading and secondary gains ($\sigma_1$ and $\sigma_2$) for $St_L > 0.5$ enables the application of the rank-1 assumption in the present study.  By using the rank-1 assumption, we focus on the leading gain and its corresponding forcing and response modes \citep{mckeon2010critical,gomez2014origin, Schmidt:JFM2017}. Here we choose the discounting parameter of $\kappa D/u_\infty=0.2$ in the resolvent analysis to analyze the unstable system. The shown gain distribution confirms that the use of  rank-1 assumption is valid for the supersonic turbulent cavity flow.

\begin{figure}
\centering
\includegraphics [width=0.5\textwidth]{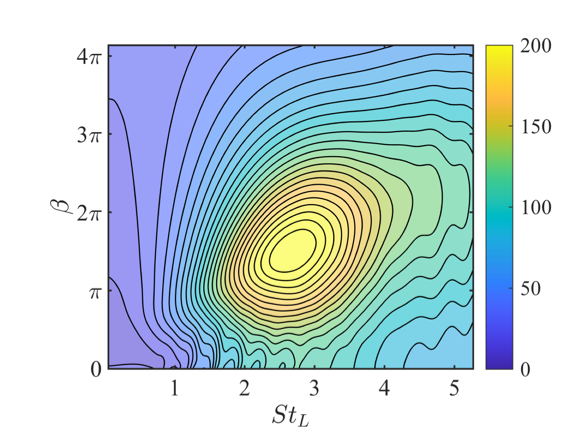}
\caption{
The leading energy amplification $\sigma_1$ for a discount parameter of $\kappa D/u_\infty=0.2$.}	\label{fig:resolvent_gain}
\end{figure}

\begin{figure}
\centering
\includegraphics[width=1.0\textwidth]{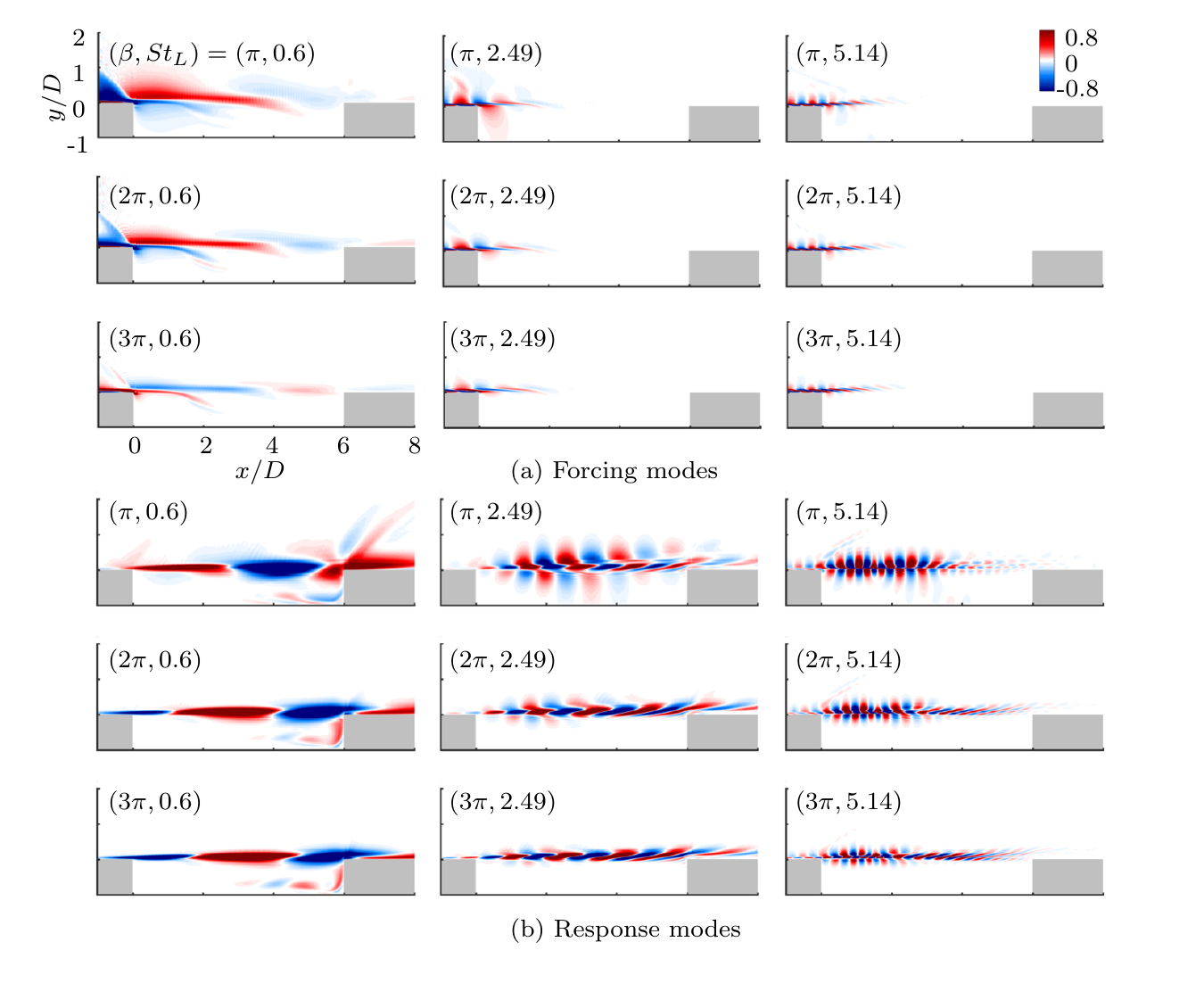}
\caption{Real streamwise component of the (a) forcing and (b) response modes at $St_L=0.6$, $2.49$ and $5.14$ with $\beta=\pi$, $2\pi$ and $3\pi$.}	
\label{fig:resolvent_mode}
\end{figure}

Figure \ref{fig:resolvent_gain} shows the leading singular value versus various combinations of spanwise wavenumber and frequency, which indicates optimal energy amplification of harmonic forcing. Larger amplification that emerges around $2<St_L<4$ and $\pi<\beta<3\pi$ is apparent in this figure. For cases of the spanwise wavenumber $\beta<\pi$, the gain distribution exhibits discrete peaks at the Rossiter mode frequencies, as they can be seen from the wavy contour lines near $\beta = 0$.   On the other hand, for spanwise wavenumber $\beta > \pi$, the gain exhibits a smoother distribution of the pseudospectral contour across the Strouhal number.

The representative forcing and response modes are presented in figure \ref{fig:resolvent_mode} for $\beta=\pi$, $2\pi$ and $3\pi$ at frequency of $St_L=0.6$, 2.49 and 5.14. The forcing and response modes are shown with the real streamwise component in figure \ref{fig:resolvent_mode}(a) and (b), respectively.  The influence of frequency and spanwise wavenumber on mode structures are similar for both forcing and response modes. As the frequency increases, the streamwise structures show a finer pattern. The structures become more compact in the transverse direction as the spanwise wavenumber increases. For the forcing modes, the dominant structures emerge at the leading edge of the cavity. It indicates the optimal spatial location for the forcing input and will be used later for the placement of unsteady control actuation. For the response modes, noticeable structures appear over shear layer region and inside the cavity.  As the frequency increases, the response mode structures become confined only over the shear-layer region.

Compression wave structures emerge at the leading and trailing edges of the cavity for the case of $(\beta, St_L)=(\pi, 0.6)$. Such wave structures become weaker and disappear for larger values of $\beta$. This change is due to the two-dimensional property of compression waves. As the spanwise wavenumber increases, the three-dimensionality suppresses the generation of two-dimensional compression waves. For the case of $St_L=2.49$, the compression waves structures disappear. For higher frequencies, the compression waves appear at the leading edge, as shown for the cases of $(\pi, 5.14)$ and $(2\pi, 5.14)$. The higher frequency forcing magnifies the obstruction effect at the leading edge, which results in the formation of compression waves in the response structures.


\section{Controlled cavity flows}
\label{sec:control}

In this section, the findings from the above resolvent and DMD analyses are used to design active flow control strategies for the attenuation of intense pressure fluctuations over the cavity.  The effectiveness of the flow control techniques is tested in full nonlinear LES. For all controlled cases considered below, the oscillatory momentum coefficient defined in (\ref{eqn:actuation}) is set to $0.02$.

\subsection{Resolvent analysis based active flow control design}
\label{sec:relevent_2_control}

Based on the knowledge gained from the DMD analysis of the instantaneous flow fields and the resolvent analysis about the time-averaged supersonic turbulent flow, we develop an active flow control strategy for effective suppression of the pressure fluctuations over the cavity.  As discussed above, the DMD analysis reveals the coherent structures that are responsible for large-scale unsteadiness over the cavity.  In a complementary manner, the resolvent analysis identifies structures that can be amplified through sustained forcing.  Here, we look for the means to amplify input perturbations over the cavity that would enhance mixing to inhibit the generation of large-scale spanwise vortical structures, especially those that are recognized by DMD as being undesirable structures.

The DMD analysis of the uncontrolled flow showed that the primary oscillation at frequency of $St_L=1.33$ is associated with the spanwise wavenumber of $\beta=\pi$.  It is the generation of the coherent structures associated with these parameters that we aim to avoid.
For these reasons, introducing forcing inputs with frequencies and spanwise wavenumbers in the vicinity of the aforementioned parameter values should be avoided.  
In fact, the goal is to force the cavity flow in a sustained manner in hopes of disrupting the formation of large undesirable structures.  These forcing inputs should remain present over the entire cavity length without spilling their energy towards the naturally energetic structures.  It is noteworthy that previous experiments by \cite{lusk2012leading} show that three-dimensional steady blowing does not last the entire length.  The key in the present study is to induce unsteady forcing inputs that persist over the cavity.  

As a guiding metric to assess the level of sustained excitation that forcing can introduce, the integrated kinetic energy along the streamwise direction from the leading response mode is considered
\begin{equation}
	{\hat E}_1(x)=\int_{-D}^{\infty} \frac{1}{2}\sigma_1^2 |\hat{\rho}_1|
	\left[ \hat{u}_1^2(x,y)+\hat{v}_1^2(x,y) \right] {\text d}y,
	\label{eqn:KEmetric}
\end{equation}
where $\sigma_1$ is the leading gain and $\hat{\rho}_1$, $\hat{u}_1$ and $\hat{v}_1$ are the corresponding density, streamwise and vertical components of the response mode, respectively.

\begin{figure}
\centering
    \includegraphics [width=0.67\textwidth]{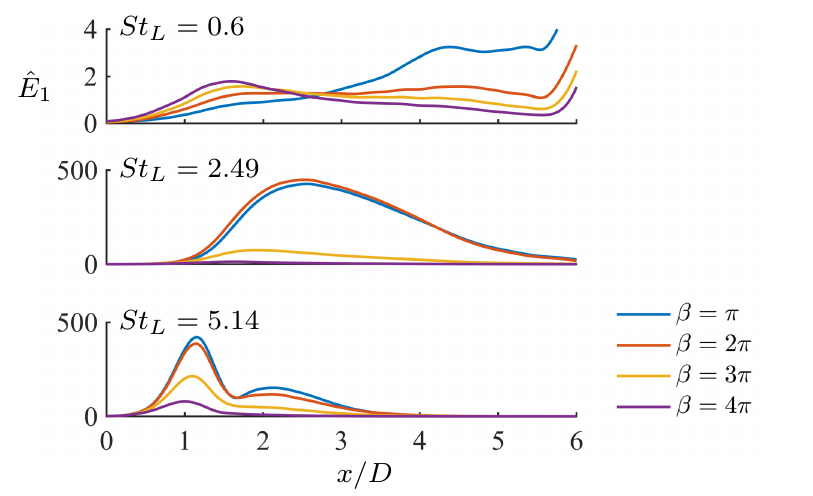}
    \caption{Responsive kinetic energy amplification profiles based on the primary response mode over the cavity at frequencies of (a) $St_L=0.6$, (b) $St_L=2.49$ and (c) $St_L=5.14$ for the spanwise wavenumbers of $\beta=\pi, 2\pi, 3\pi$ and $4\pi$.}
    \label{metrics}
\end{figure}

The primary kinetic energy amplification profiles $\hat{E}_1(x)$ over the cavity for representative cases at frequencies of $St_L=0.6$, $2.49$ and $5.14$ and spanwise wavenumbers of $\beta=\pi$, $2\pi$, $3\pi$ and $4\pi$ are presented in figure \ref{metrics}. 
For the cases at $St_L=0.6$, the energy amplification over the cavity maintains a low magnitude, reaching only up to $\hat{E}_1 = 4$ compared to those cases from the other two frequencies. 
The distribution changes over the cavity as the spanwise wavenumber increases. 
It is observed that primary energy amplifications are larger for higher $\beta$ in the front half of the cavity but reverses past $x/D \approx 3$.

Next, we turn our attention to the primary energy amplification at a higher frequency of $St_L = 2.49$. Compared to the previous cases at $St_L=0.6$, the energy amplifications here are two orders of magnitude larger over the cavity.  
The perturbations with spanwise wavenumbers of $\beta = \pi$ and $2\pi$ are especially noteworthy.  
For these cases, the primary kinetic energy amplification remains high over the entire length of the cavity, with its maximum value attain around the middle of the cavity.  
For spanwise wavenumbers of $\beta=3\pi$ and $4\pi$, the magnitudes of $\hat{E}_1$ are significantly lower.  
The sustained high-level energy amplification is attractive as a means to perform active flow control if such modal response is desirable over the entire cavity length.  Furthermore, since the frequency of $St_L = 2.49$ is higher than what is identified from DMD analysis, the frequency and spanwise wavenumbers appear to be viable parameter choices for effective flow control.

For cases at the frequency of $St_L=5.14$, the responsive energy significantly amplifies in the front of the cavity. 
It dramatically decreases around  $x/D\approx 1$ and retains at low magnitude in the rear part of the cavity. 
This distribution of the responsive kinetic energy, losing its authority in the rear part of the cavity, is considered an adverse control effect. Other large coherent structures may emerge when the responsive kinetic energy is weak to sustainable forcing. For this reason, it appears that such high-frequency forcing will be less effective for the suppression of fluctuations in open cavity flows.

Based on the understanding of the baseline cavity flow gained from DMD and resolvent analyses as well as the above discussions, we propose the following flow control strategy for cavity flow.  
First, avoid introducing perturbations at forcing frequencies associated with detrimental large-scale unsteadiness, as identified from DMD analysis of the baseline flow (here, $St_L \approx 1.33$).
Second, select combinations of the actuation frequency and spanwise wavenumber that can sustain the amplification of small-scale structures (streamwise vortices) over the entire cavity length.  
For this flow, we have used the kinetic energy distribution from the primary response mode and found that combinations of forcing frequency $St_L \approx 2.49$ with spanwise wavenumbers $\beta \approx \pi$ to $3\pi$ are promising candidates for achieving effective suppression of cavity flow fluctuations.

\subsection{Assessments of resolvent analysis based cavity flow control} 
\label{sec:cc}

Unsteady actuation in LES is introduced with the forcing frequency $St_c$ and spanwise wavenumber $\beta_c$ suggested by the resolvent analysis.  
Because practical forcing cannot be introduced globally, local forcing is imposed along the cavity leading edge, which is where the forcing modes are particularly concentrated.
In the previous discussion, we identified that forcing frequency of $St_L=2.49$ with spanwise wavenumbers between $\pi$ to $3\pi$ show a sustained responsive kinetic energy over the cavity from the resolvent analysis. 
LES are performed for these parameters and some representative parameters to assess the effectiveness of the actuation setup.
We in particular consider forcing frequency and spanwise wavenumber in the range of $0.6\le St_c \le 5.14$ and $\pi \le \beta_c \le 4\pi$ for the following validation LES cases.

\begin{figure}
\centering
\includegraphics[width=\textwidth]{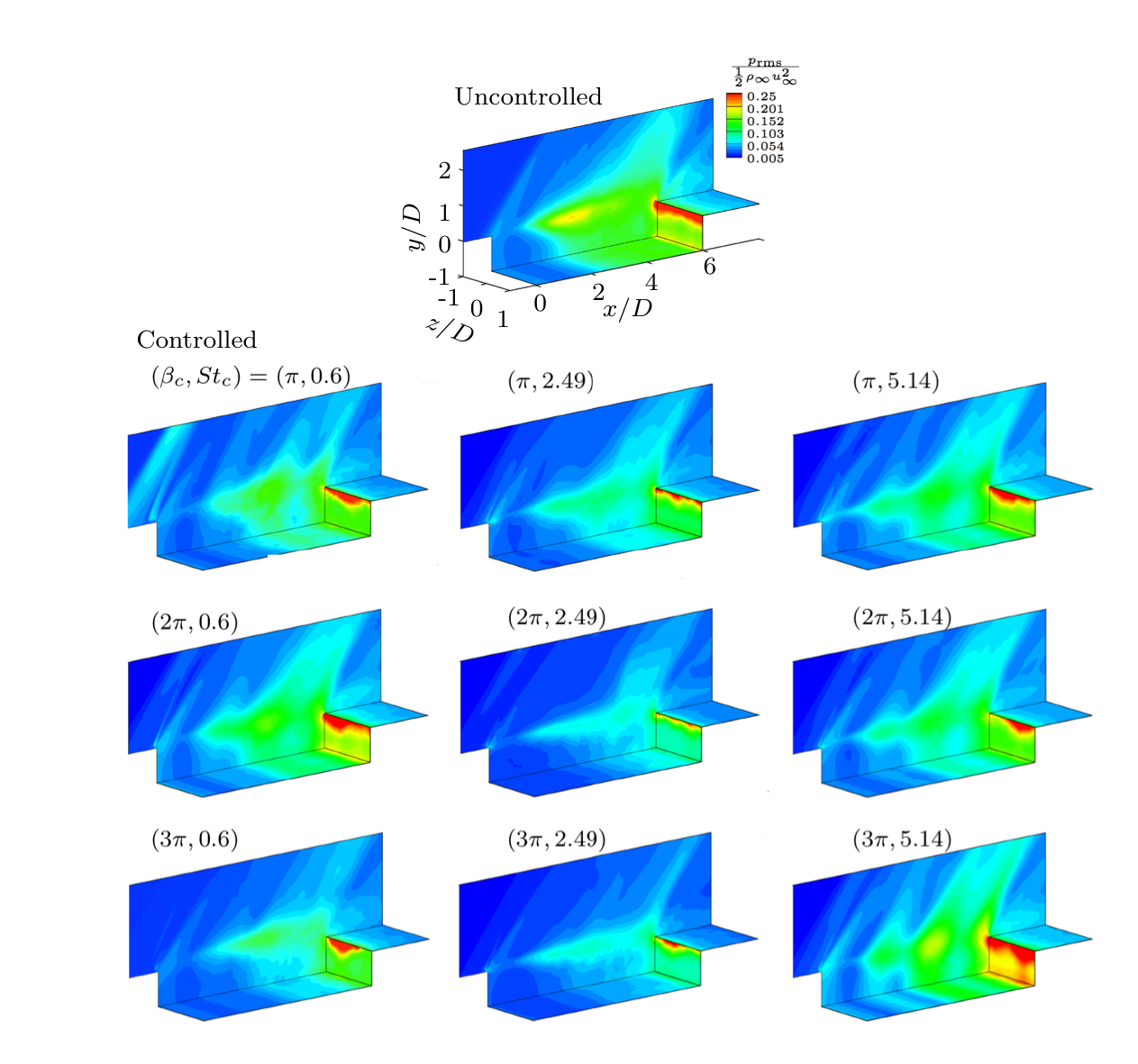}
\caption{The pressure RMS $p_\text{rms}/(\frac{1}{2}\rho_\infty u^2_\infty)$ over the cavity surfaces and the $x-y$ plane at $z/D=-1$ for the uncontrolled (top) and controlled cases.}	
\label{fig:cp_rms} 
\end{figure}

The pressure fluctuations obtained from LES of controlled flows are examined over the cavity surfaces and the shear layer.  
Intense pressure oscillations appear over these regions in the uncontrolled cavity flows.
Nine representative controlled cases are investigated, including the controlled cases with the  parameters that are identified to provide sustained forcing over the cavity by resolvent analysis. The pressure fluctuations (RMS) for the controlled cases are shown in figure \ref{fig:cp_rms}.  Compared to the pressure RMS distribution for the uncontrolled flow, we observe that some of the controlled cases show significant attenuation of the fluctuations over the shear layer and on the cavity walls.

For the control cases with $St_c=0.6$, the pressure fluctuations are reduced over the shear-layer region and along the cavity walls compared to the uncontrolled cavity flow. 
Along the cavity floor, the area experiencing large pressure fluctuations are limited to the latter third of the cavity, which is reduced in size from the uncontrolled case.
Above the shear layer, actuation is able to modify the flow field to reduce the spatial extent over which large fluctuations appear, but the unsteady shock-induced fluctuations remain strong over the rear half of the cavity in the case of forcing spanwise wavenumbers of $\beta_c = \pi$ and $2\pi$.

Next, the controlled flows with a forcing frequency of $St_c=2.49$ are considered, which is the frequency hypothesized to perform well from resolvent analysis.  All of the forcing spanwise wavenumbers of $\beta_c = \pi$, $2\pi$, and $3\pi$ shown in figure \ref{fig:cp_rms} significantly reduce the pressure fluctuations over the cavity walls.  The oscillations above the shear layer are also reduced.   
Noteworthy here are the cases where the spanwise wavenumbers of the forcing input are $2\pi$ and $3\pi$.  For these cases, the unsteady forcing is able to reduce the spatial extent of large fluctuations.  Due to the stabilizing effect in the shear layer and weakened impingement of flow structures, we observe remarkable reduction in pressure fluctuations above $50\%$ along the aft wall of the cavity (see later summary, as we summarized later in table \ref{tab:kd}).  Compared to the baseline and the other controlled cases, high level of pressure fluctuations are experienced only along the trailing edge of the cavity.
The exceptional reduction of pressure fluctuations over the cavity is achieved with the choice of control parameters identified from the resolvent (response mode) based kinetic energy based metric (\ref{eqn:KEmetric}).  This implies that the input-output relationship captured by resolvent analysis for the turbulent cavity flow indeed can point to the effective set of control parameters with significantly less computational resources than what is required by an uninformed parametric LES study.

Forcing at a frequency of $St_c=5.14$ is also considered, which is higher than the frequency identified from resolvent analysis to be effective.  For these controlled cases, the pressure fluctuations over the shear layer are higher in magnitude compared to the cases of $St_c=2.49$, as shown in figure \ref{fig:cp_rms}. These results are expected as the forcing to enhance mixing across the shear layer is not as sustained as in the case of $St_c=2.49$, which is indicated by figure \ref{metrics}. In fact, the unsteady control induces the appearance of oblique shocks for all spanwise wavenumber cases shown in figure \ref{fig:cp_rms}, which renders the control counter-productive.  We also observe strong impingement of vortical structures for the spanwise wavenumber of $3\pi$, which increases the pressure fluctuations on the entire aft wall above that of the uncontrolled case.

\begin{figure}
    \centering
    \includegraphics [width=0.9\textwidth]{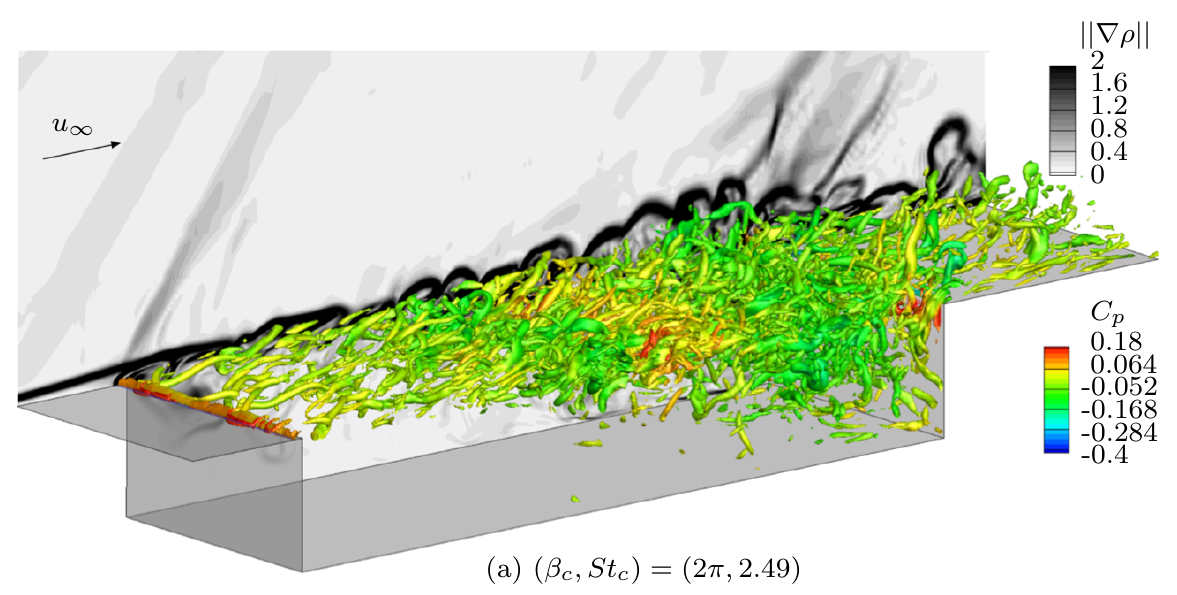}\\
    \includegraphics [width=0.9\textwidth]{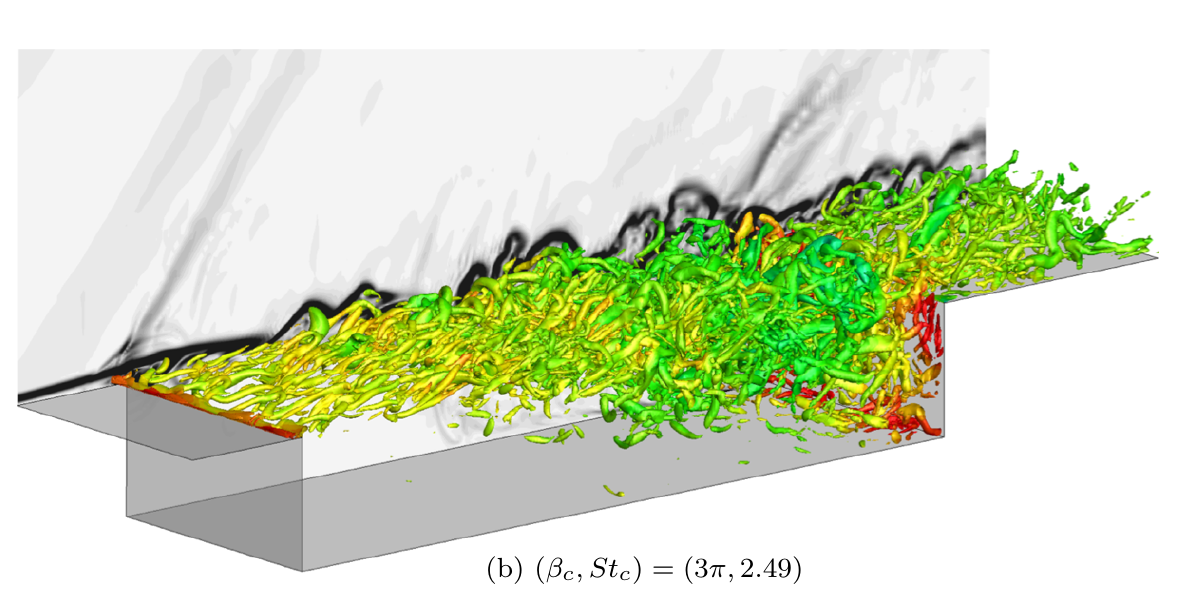}\\
    \includegraphics [width=0.9\textwidth]{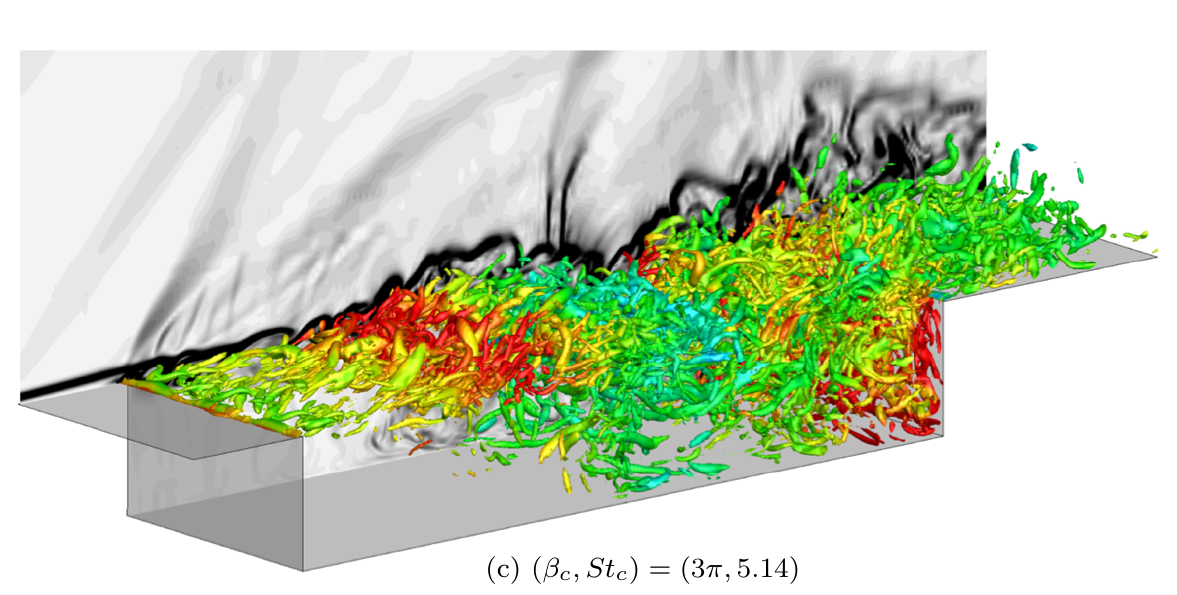}
    \caption{
    Instantaneous controlled cavity flow with (a) $(\beta_c, St_c)=(2\pi,2.49)$, (b) $(3\pi,2.49)$ and (c) $(3\pi, 5.14)$, visualized with the $Q$-criterion $Q(D/u_\infty)^2=10$ colored by $C_p=(p-p_\infty)/(\frac{1}{2}\rho_\infty u^2_\infty)$ at $M_\infty=1.4$. Numerical schlieren is shown in the background in gray scale.
    }
    \label{fig:c_Q}
\end{figure}

Looking closer at the flow fields for some of the cases presented in figure \ref{fig:cp_rms}, we visualize in figure \ref{fig:c_Q} the instantaneous vortical structures for effective controlled cases with $(\beta_c, St_c)=(2\pi,2.49)$ and $(3\pi,2.49)$ as well as an ineffective controlled case with $(3\pi, 5.14)$.  
Shown in this figure are the iso-surfaces of the $Q$-criterion colored by the pressure coefficient.  
For the case of $(\beta_c, St_c)=(2\pi,2.49)$ shown in figure \ref{fig:c_Q}(a), small streamwise vortices are generated from the leading edge and propagate downstream, which gradually spread over the shear layer. Large-scale spanwise shear layer roll-ups disappear in the flow field compared to the uncontrolled cavity flow (see figure \ref{fig: baseline_vortices}).
The modification of the flow field produced by the unsteady actuation greatly attenuates the large-scale flow structure impingement on the aft-wall by breaking up large structures and reduces the fluctuation over the cavity.  
The aft wall only experiences small-scale vortical structures hitting the wall in an incoherent manner.  
As the kinetic energy profile (\ref{eqn:KEmetric}) foreshadows, the mixing and breakup of the large-scale spanwise vortices are sustained over the entire cavity length and yields an effective control approach to reduce the pressure fluctuations.

For the controlled case with $(\beta_c, St_c)=(3\pi, 2.49)$ presented in figure \ref{fig:c_Q}(b), the flow structure appears similar to those from the case of $(\beta_c, St_c)=(2\pi,2.49)$.  With the choice of a high wavenumber, the structures are generated by the unsteady actuation from the cavity leading edge in a closely packed manner.  
These streamwise vortical structures appear to cancel the vortical influence from each other due to their close proximity and cannot remain sufficiently influential to suppress the spanwise instability from appearing as effective as the case with $(\beta_c, St_c)=(2\pi, 2.49)$.  
This leads to a corresponding larger fluctuations over the cavity and local thickening of the shear layer, as visualized by the wider range of pressure values in figure \ref{fig:c_Q}(b).

For the case of $(\beta_c, St_c)=(3\pi,5.14)$ shown in figure \ref{fig:c_Q}(c), the streamwise vortices generated from unsteady actuation are truncated to become shorter than what are observed for the lower frequency actuation. 
By the middle of the cavity, the shear layer rolls up into a spanwise vortical coherent structure with a size comparable to the cavity depth that penetrates both into the freestream and the cavity.
The local obstructions of the incoming flow by these large-scale vortical structures lead to the appearance of strong compression waves, adding to the increased pressure fluctuations over the shear layer and along the cavity walls.
Compared to the other controlled cases, strong acoustic waves emitted from the leading edge of the cavity are observable.

\begin{figure}
\includegraphics [width=\textwidth]{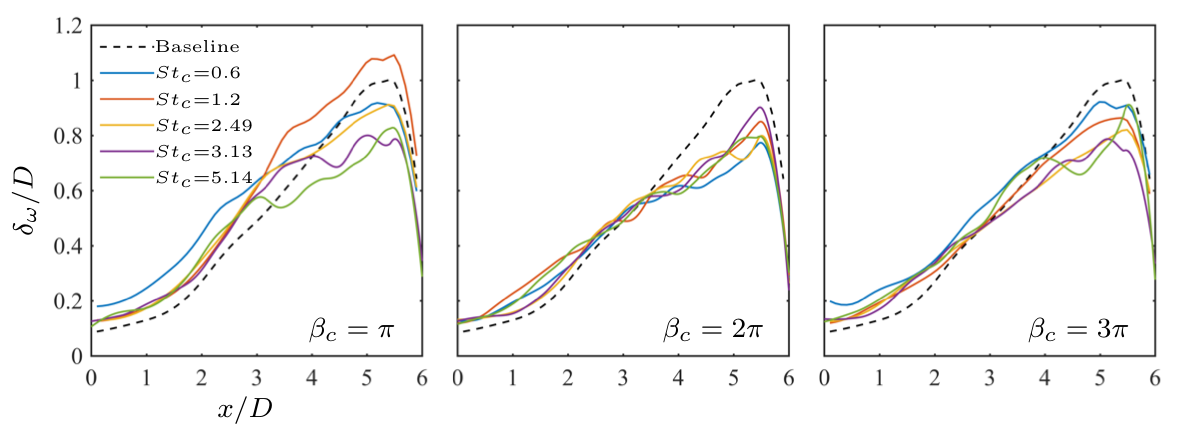}
\caption{Vorticity thickness for the uncontrolled and controlled cases.}	
\label{fig:vt}
\end{figure}

It is observed above that the shear-layer thickness is affected by the control input and influences the level of pressure fluctuations.  
For this reason, the vorticity thickness 
\begin{equation}\delta_\omega = \frac{u_\infty}{(\partial \bar{u}/\partial y)_{\max}}
\end{equation} 
is considered to quantify the direct influence of the unsteady actuation on cavity flow.  Here, $\bar{u}$ is the spanwise and time-averaged streamwise velocity.
Figure \ref{fig:vt} shows the vorticity thickness of the shear layer from controlled cases at spanwise wavenumbers of $\beta_c=\pi, 2\pi$, and $3\pi$. For the controlled cases, the vorticity thickness becomes thicker at the leading edge compared to the uncontrolled case.  This local thickening of the shear layer at the leading edge reduces its receptivity to the acoustic disturbances and stabilizes the shear layer as it advects downstream. For the cases of $\beta_c=2\pi$, a significant reduction of vorticity thickness is observed in the rear part of the cavity. Note that the sudden decrease of the vorticity thickness near the trailing edge is due to the effect of impingement on the cavity trailing edge.

The magnitude of pressure fluctuations inside the cavity is coupled with the pressure fluctuations occurring above the shear layer region. We quantify the control effectiveness by integrating the pressure fluctuations over the floor and aft-wall of the cavity using definition (\ref{eqn:assess}). The unsteady control cases in general achieve significant reductions of pressure fluctuations compared to the uncontrolled case.  Particularly noteworthy is the control case with $(\beta_c, St_c)=(2\pi, 2.49)$ which achieves $52.1\%$ reduction in the pressure fluctuations, as summarized in table \ref{tab:kd}. Additionally, the present unsteady control outperforms our previous control study using steady actuation \citep{sun2018effects}. The trends summarized in table \ref{tab:kd} are consistent with the predictive metric for control design, which validates the conceptual approach of how to use the results of the resolvent analysis to determine an effective unsteady flow control setup.

\begin{table}
  \begin{center}
\def~{\hphantom{0}}
  \begin{tabu}to 0.85\textwidth{l X[r] X[r] X[r] X[r]}
   Cases    & $\beta_c$ & $St_c$ & $\tilde{p}_\text{rms}$ & $\Delta \tilde{p}_\text{rms}$ \\ [2pt] \hline 
  \\ [-2pt]
  Baseline &  -  & - & 1.369   & - \\[2pt] 
  \hline\\
  [-3pt]
 								     & $ \pi$  & 0.6   &0.983    & -28.2$\%$  \\
                                     & $ \pi$  & 1.2   &0.847    & -38.1$\%$  \\ 
                                     & $ \pi$  & 2.49  &0.851    & -37.8$\%$  \\
                                     & $ \pi$  & 3.13  &0.939    & -31.4$\%$  \\ 
                                     & $ \pi$  & 5.14  &1.129    & -17.5$\%$  \\
                                     [5pt]
                                     & $2\pi$  & 0.6   &1.249    &  -8.8$\%$  \\
                                     & $2\pi$  & 1.2   &0.912    & -33.4$\%$  \\
   \multirow{3}{*}{Unsteady control} & $2\pi$  & 2.49  &0.656    & -52.1$\%$  \\ 
                                     & $2\pi$  & 3.13  &0.960    & -29.9$\%$  \\ 
                                     & $2\pi$  & 5.14  &1.052    & -23.2$\%$  \\[5pt]                                      
                                     & $3\pi$  & 0.6   & 0.821   & -40.0$\%$  \\
                                     & $3\pi$  & 1.2   & 0.796   & -41.9$\%$  \\
                                     & $3\pi$  & 2.49  & 0.667   & -51.3$\%$  \\ 
                                     & $3\pi$  & 3.13  & 1.076   & -21.4$\%$  \\
                                     & $3\pi$  & 5.14  & 1.435   &  +4.8$\%$  \\
                                     [2pt]
                                     & $4\pi$  & 2.49  & 0.881   & -35.6$\%$  \\ [2pt]
  \hline
  \\ [-3pt]
  Steady control \citep{sun2018effects}   & $3\pi$ & 0  & 0.981   & -28.3$\%$  \\ [2pt] \hline
  \end{tabu}
  \caption{
  Summary of flow control cases with unsteady actuation.}
  \label{tab:kd}
  \end{center}
\end{table}

\subsection{Dynamic mode decomposition of the controlled flows}
\label{sec: dmd}

\begin{figure}
\centering
\includegraphics [width=\textwidth,trim={0 0cm 0 0cm},clip]{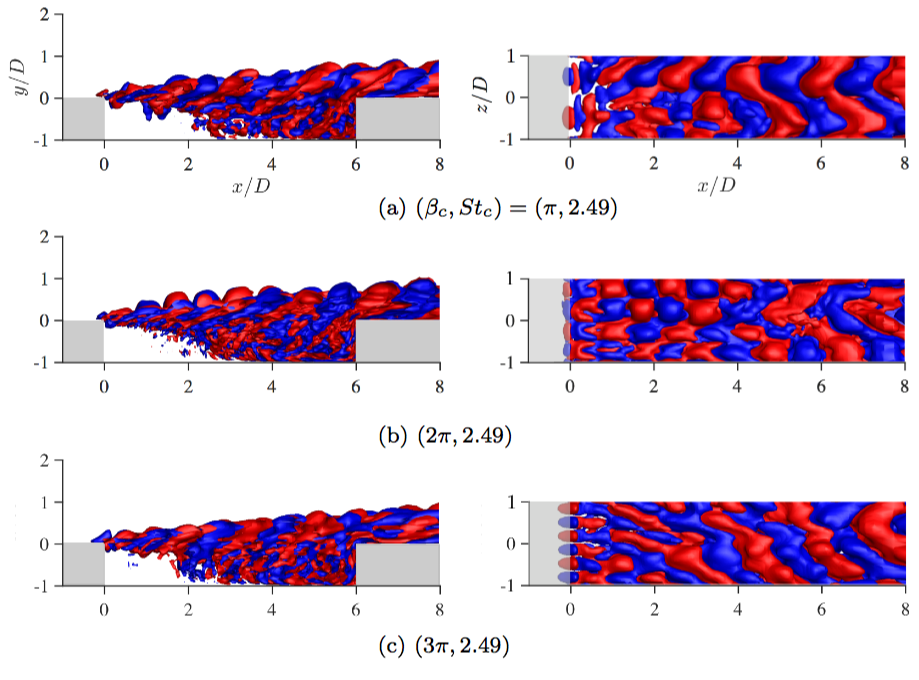}
\caption{Side and top views of the real spanwise velocity of DMD modes for the control cases. Iso-surface values of streamwise velocity are $0.1$ (red) and $-0.1$ (blue).}	
\label{fig:dmd_controlled}
\end{figure}

Let us further assess the influence of unsteady actuation on the turbulent cavity flow using DMD analysis.  We extract the coherent structures produced at the forcing frequency to examine the persistence of forcing input (perturbation) from the leading edge of the cavity.  Here, DMD analysis is performed on the controlled cavity flows for the forcing frequency of $St_c=2.49$ with spanwise wavenumbers of $\beta_c = \pi$, $2\pi$ and $3\pi$ to uncover the key flow response around the optimal control case.  Presented in figure \ref{fig:dmd_controlled} are the spatial structures of the spanwise velocity associated with the forcing frequency of $St_c=2.49$.  By examining the spatial evolution of the DMD modes over the cavity, we can understand the mechanism by which the forcing input prevents the undesirable large-scale vortex roll-up from appearing over the cavity.

Consider the controlled flow case of $\beta_c=\pi$, shown in figure \ref{fig:dmd_controlled}(a).  This forcing wavenumber is the same as the dominant spanwise wavenumber observed in figure \ref{fig:dmd_mode} for the uncontrolled flow.   From the top view in figure \ref{fig:dmd_controlled}(a), we notice that the flow structures essentially maintain the spanwise wavenumber of $\beta=\pi$, except for some regions over $1 \lesssim x/D \lesssim 3$ due to nonlinear interactions.  The fact that $\beta=\pi$ remains the primary wavenumber indicates that the control enhances rather than attenuates the undesirable fluctuations.

For the case of controlled flow with $\beta_c=2 \pi$, shown in figure \ref{fig:dmd_controlled}(b), we observe that the flow structures predominantly possess the spanwise wavenumber of $\beta=2\pi$ almost over the entire cavity ($x/D \lesssim 6$).  This means that the actuation remains effective over the whole length of the cavity, providing no opportunity for the undesirable large-scale unsteady vortical structure, with $\beta=\pi$ to emerge.  As discussed earlier, the flow control input for this flow produces a range of small-scale structures, reducing the spatial fluctuation level over the shear layer.  This modification of the flow in turn avoids the generation of unsteady oblique shock waves over the cavity.

For the higher control input wavenumber of $\beta_c=3\pi$, we observe the finer structures introduced from the leading edge of the cavity, as seen in figure \ref{fig:dmd_controlled}(c). As the flow convects downstream, the spanwise vortical structures start merging past $x/D \approx 2$.  Once this spanwise merging process takes place, the flow relaxes back to the large-scale structures with $\beta=\pi$ and leads to the weakening of control effects.  However, full relaxation does not appear until $x/D \approx 3$, which allows the flow to resist the formation of the large-scale structures.  From the insights gain through the DMD-based analysis, we can assess the spatial extent over which the actuation input remains effective.  The present DMD analysis confirms that the choice of $(\beta_c, St_c) = (2\pi, 2.49)$ is indeed a good choice for attenuating fluctuations generated in the present supersonic turbulent cavity flow.

\section{Conclusions}
\label{sec:summary}

We examined a supersonic turbulent flow over a spanwise-periodic rectangular cavity at a Mach number of 1.4 and a cavity-depth-based Reynolds number of $10,000$.  The considered cavity has a length-to-depth ratio of $6$ and a spanwise periodic extent of $2$.  Large-eddy simulations and dynamic mode decomposition of the unsteady turbulent cavity flow were performed to gain detailed insights on the source of large pressure fluctuations.  The formations of large-scale vortical structures associated with Rossiter modes II and IV were found to be responsible for the generation of high-amplitude unsteadiness for this case.  To attenuate large pressure fluctuations, it was deduced that the emergence of these Rossiter modes should be hindered.  

In an effort to develop an effective active flow control technique to reduce the level of pressure fluctuations over the cavity, resolvent analysis was performed about the time-averaged supersonic turbulent cavity flow.  The resolvent analysis uncovered the dominant input-output characteristics with a leading pair of forcing and response modes and the gain that represents the amount of amplification between the two modes for a combination of forcing frequency and spanwise wavenumber.  The results from resolvent analysis revealed significant energy amplification can be achieved for the principal mode pairs for the frequencies of $1.5\lesssim St_L \lesssim 4$ and spanwise wavenumbers of $\pi \lesssim \beta \lesssim 3\pi$.  The response modes showed spatial oscillations over the shear layer with the corresponding forcing mode exhibiting high amplitude input near the leading edge of the cavity.  These findings suggested that active flow control may effectively alter the flow by introducing the three-dimensional (spanwise-varying) forcing input along the leading edge in the identified range of frequency and spanwise wavenumber.

To further develop a reliable control guideline, we sought control cases that introduce actuation input that is sustained over the entire length of the cavity.  This translates to ensuring that the streamwise vortices introduced from the unsteady actuation input can effectively suppress the formation of the dominant spanwise vortical structures over the cavity.  In order to find such an optimal control setting, we assess the kinetic energy distribution from the primary response mode over the cavity.  The combination of the forcing frequency and spanwise wavenumber, $(St_L, \beta) = (2.49, 2\pi)$, that provided the highly amplified and sustained distribution of the kinetic energy over the cavity was determined to be the optimal candidate for attenuating the pressure oscillations.  

Large-eddy simulations of representative unsteady flow control cases with $C_\mu=0.02$ verified that the resolvent-analysis-based control guideline is indeed effective in determining the appropriate flow control input for attenuating the unsteadiness in cavity flow.  The simulations introduced unsteady blowing with spanwise variation along the cavity leading edge.  The simulated flows showed that the optimal flow control candidate achieved a remarkable rms pressure reduction along the cavity walls of $52\%$.  The control input reduced the pressure fluctuations by hindering the emergence of large spanwise vortices and by eliminating the obstructions responsible for creating oblique shock waves over the cavity.  The controlled flow was further studied with DMD analysis, revealing that modal structures associated with Rossiter modes were eliminated with actuation.  It should be further noted that the present unsteady control technique was able to achieve reduction of pressure fluctuations by nearly double of what had been achieved by steady mass injection in our previous study based on biglobal stability analysis \citep{sun2018effects}.

The present study demonstrated that the physical insights gained from resolvent analysis of turbulent supersonic flow can be used to develop an effective flow control approach.  This formulation provides an alternative to the trial-and-error based optimization to determine an effective flow control strategy, which requires considerable efforts in terms of computational and experimental settings.  While the current work was able to determine the optimal control setup, actuation parameters in the vicinity of the identified parameters should also be examined to ensure the optimality of the control effectiveness.  The present approach should provide pathways to perform flow control for which the added perturbations need to remain effective in modifying the flow in a sustained fashion.  Moreover, extensions of resolvent-analysis-based control to turbulent flows at higher Reynolds number should also be possible with emerging resolvent-based techniques \citep{Ribeiro:PRF20, Yeh:PRF20} that incorporate numerical algorithms that accelerate computations and provide relief on computational memory.

\section*{Acknowledgments}

The authors gratefully acknowledge the support by the U.S. Air Force Office of Scientific Research (Award number: FA9550-17-1-0380; Program Managers: Dr. Douglas R. Smith and Dr. Gregg Abate).


\begin{thebibliography}{60}
\expandafter\ifx\csname natexlab\endcsname\relax\def\natexlab#1{#1}\fi
\def\au#1{#1} \def\ed#1{#1} \def\yr#1{#1}\def\at#1{#1}\def\jt#1{\textit{#1}}
  \def\bt#1{#1}\def\bvol#1{\textbf{#1}} \def\vol#1{#1} \def\pg#1{#1}
  \def\publ#1{#1}\def\arxiv#1{#1}\def\org#1{#1}\def\st#1{\textit{#1}}

\bibitem[Bechara {\em et~al.\/}(1994)Bechara, Bailly, Lafon \&
  Candel]{bechara1994stochastic}
{\sc \au{Bechara, W.}, \au{Bailly, C.}, \au{Lafon, P.} \& \au{Candel, S.~M.}}
  \yr{1994}  \at{Stochastic approach to noise modeling for free turbulent
  flows}.  \jt{AIAA J.}  \bvol{32}~(3),  \pg{455--463}.

\bibitem[Beresh {\em et~al.\/}(2016)Beresh, Wagner \& Casper]{Beresh:JFM2016}
{\sc \au{Beresh, S.~J.}, \au{Wagner, J.~L.} \& \au{Casper, K.~M.}} \yr{2016}
  \at{Compressibility effects in the shear layer over a rectangular cavity}.
  \jt{J. Fluid Mech.}  \bvol{808},  \pg{116--152}.

\bibitem[Br\`es \& Colonius(2008)]{bres2008three}
{\sc \au{Br\`es, G.~A.} \& \au{Colonius, T.}} \yr{2008}  \at{Three-dimensional
  instabilities in compressible flow over open cavities}.  \jt{J. Fluid Mech.}
  \bvol{599},  \pg{309--339}.

\bibitem[Br\`es {\em et~al.\/}(2017)Br\`es, Ham, Nichols \& Lele]{Bres:AIAAJ17}
{\sc \au{Br\`es, G.~A.}, \au{Ham, F.~E.}, \au{Nichols, J.~W.} \& \au{Lele,
  S.~K.}} \yr{2017}  \at{Unstructured large-eddy simulations of supersonic
  jets}.  \jt{AIAA J.}  \bvol{55}~(4),  \pg{1164--1184}.

\bibitem[Casper {\em et~al.\/}(2018)Casper, Wagner, Beresh, Spillers, Henfling
  \& Dechant]{casper2018spatial}
{\sc \au{Casper, K.~M.}, \au{Wagner, J.~L.}, \au{Beresh, S.~J.}, \au{Spillers,
  R.~W.}, \au{Henfling, J.~F.} \& \au{Dechant, L.~J.}} \yr{2018}  \at{Spatial
  distribution of pressure resonance in compressible cavity flow}.  \jt{J.
  Fluid Mech.}  \bvol{848},  \pg{660--675}.

\bibitem[Cattafesta \& Sheplak(2011)]{cattafesta2011actuators}
{\sc \au{Cattafesta, L.~N.} \& \au{Sheplak, M.}} \yr{2011}  \at{Actuators for
  active flow control}.  \jt{Annu. Rev. Fluid Mech.}  \bvol{43},
  \pg{247--272}.

\bibitem[Cattafesta {\em et~al.\/}(2008)Cattafesta, Song, Williams, Rowley \&
  Alvi]{cattafesta2008active}
{\sc \au{Cattafesta, L.~N.}, \au{Song, Q.}, \au{Williams, D.~R.}, \au{Rowley,
  C.~W.} \& \au{Alvi, F.~S.}} \yr{2008}  \at{Active control of flow-induced
  cavity oscillations}.  \jt{Prog. Aerosp. Sci.}  \bvol{44}~(7-8),
  \pg{479--502}.

\bibitem[Chu(1965)]{chu1965energy}
{\sc \au{Chu, B.-T.}} \yr{1965}  \at{On the energy transfer to small
  disturbances in fluid flow (part {I})}.  \jt{Acta Mechanica}  \bvol{1}~(3),
  \pg{215--234}.

\bibitem[Colonius(2001)]{colonius2001overview}
{\sc \au{Colonius, T.}} \yr{2001}  \bt{An overview of simulation, modeling, and
  active control of flow/acoustic resonance in open cavities}. {AIAA} Paper
  2001-0076.

\bibitem[De~Vicente {\em et~al.\/}(2014)De~Vicente, Basley, Meseguer-Garrido,
  Soria \& Theofilis]{de2014three}
{\sc \au{De~Vicente, J.}, \au{Basley, J.}, \au{Meseguer-Garrido, F.},
  \au{Soria, J.} \& \au{Theofilis, V.}} \yr{2014}  \at{Three-dimensional
  instabilities over a rectangular open cavity: from linear stability analysis
  to experimentation}.  \jt{J. Fluid Mech.}  \bvol{748},  \pg{189--220}.

\bibitem[Elimelech {\em et~al.\/}(2011)Elimelech, Vasile \&
  Amitay]{elimelech2011secondary}
{\sc \au{Elimelech, Y.}, \au{Vasile, J.} \& \au{Amitay, M.}} \yr{2011}
  \at{Secondary flow structures due to interaction between a finite-span
  synthetic jet and a 3-{D} cross flow}.  \jt{Phys. Fluids}  \bvol{23}~(9),
  \pg{094104}.

\bibitem[Faure {\em et~al.\/}(2007)Faure, Adrianos, Lusseyran \&
  Pastur]{faure2007visualizations}
{\sc \au{Faure, T.~M.}, \au{Adrianos, P.}, \au{Lusseyran, F.} \& \au{Pastur,
  L.}} \yr{2007}  \at{Visualizations of the flow inside an open cavity at
  medium range {R}eynolds numbers}.  \jt{Exp. Fluids}  \bvol{42}~(2),
  \pg{169--184}.

\bibitem[Faure {\em et~al.\/}(2009)Faure, Pastur, Lusseyran, Fraigneau \&
  Bisch]{Faure:EF2009}
{\sc \au{Faure, T.~M.}, \au{Pastur, L.}, \au{Lusseyran, F.}, \au{Fraigneau, Y.}
  \& \au{Bisch, D.}} \yr{2009}  \at{Three-dimensional centrifugal instabilities
  development inside a parallelepipedic open cavity of various shape}.
  \jt{Exp. Fluids}  \bvol{47}~(3),  \pg{395--410}.

\bibitem[George {\em et~al.\/}(2015)George, Ukeiley, Cattafesta \&
  Taira]{george2015control}
{\sc \au{George, B.}, \au{Ukeiley, L.~S.}, \au{Cattafesta, L.~N.} \& \au{Taira,
  K.}} \yr{2015}  \bt{Control of three-dimensional cavity flow using
  leading-edge slot blowing}. {AIAA} Paper 2015-1059.

\bibitem[G{\'o}mez {\em et~al.\/}(2014)G{\'o}mez, Blackburn, Rudman, McKeon,
  Luhar, Moarref \& Sharma]{gomez2014origin}
{\sc \au{G{\'o}mez, F.}, \au{Blackburn, H.~M.}, \au{Rudman, M.}, \au{McKeon,
  B.~J.}, \au{Luhar, M.}, \au{Moarref, R.} \& \au{Sharma, A.~S.}} \yr{2014}
  \at{On the origin of frequency sparsity in direct numerical simulations of
  turbulent pipe flow}.  \jt{Phys. Fluids}  \bvol{26}~(10),  \pg{101703}.

\bibitem[Heller {\em et~al.\/}(1971)Heller, Holmes \& Covert]{heller1971flow}
{\sc \au{Heller, H.~H.}, \au{Holmes, D.~G.} \& \au{Covert, E.~E.}} \yr{1971}
  \at{Flow-induced pressure oscillations in shallow cavities}.  \jt{J. Sound
  Vib.}  \bvol{18}~(4),  \pg{545--553}.

\bibitem[Jovanovi\'{c}(2004)]{jovanovic2004modeling}
{\sc \au{Jovanovi\'{c}, M.~R.}} \yr{2004}  \at{Modeling, analysis, and control
  of spatially distributed systems}. PhD thesis, University of California,
  Santa Barbara.

\bibitem[Jovanovi\'{c} \& Bamieh(2005)]{jovanovic2005componentwise}
{\sc \au{Jovanovi\'{c}, M.~R.} \& \au{Bamieh, B.}} \yr{2005}  \at{Componentwise
  energy amplification in channel flows}.  \jt{J. Fluid Mech.}  \bvol{534},
  \pg{145--183}.

\bibitem[Khalighi {\em et~al.\/}(2011)Khalighi, Ham, Moin, Lele \&
  Schlinker]{Khalighi:ASME2011}
{\sc \au{Khalighi, Y.}, \au{Ham, F.}, \au{Moin, P.}, \au{Lele, S.~K.} \&
  \au{Schlinker, R.~H.}} \yr{2011} Noise prediction of pressure-mismatched jets
  using unstructured large eddy simulation.  \bt{In {\em ASME 2011 Turbo Expo:
  Turbine Technical Conference and Exposition\/}},  \pg{pp. 381--387}.

\bibitem[Krishnamurty(1955)]{krishnamurty1955acoustic}
{\sc \au{Krishnamurty, K.}} \yr{1955}  \bt{Acoustic radiation from
  two-dimensional rectangular cutouts in aerodynamic surfaces}. {\em Tech.
  Rep.\/} 3487.  \org{N.A.C.A. Tech. Note}.

\bibitem[Larchev{\^e}que {\em et~al.\/}(2007)Larchev{\^e}que, Sagaut \&
  Labb{\'e}]{larcheveque2007large}
{\sc \au{Larchev{\^e}que, L.}, \au{Sagaut, P.} \& \au{Labb{\'e}, O.}} \yr{2007}
   \at{Large-eddy simulation of a subsonic cavity flow including asymmetric
  three-dimensional effects}.  \jt{J. Fluid Mech.}  \bvol{577},  \pg{105--126}.

\bibitem[Lawson \& Barakos(2011)]{lawson2011review}
{\sc \au{Lawson, S.~J.} \& \au{Barakos, G.~N.}} \yr{2011}  \at{Review of
  numerical simulations for high-speed, turbulent cavity flows}.  \jt{Pro.
  Aerosp. Sci.}  \bvol{47}~(3),  \pg{186--216}.

\bibitem[Leclercq {\em et~al.\/}(2019)Leclercq, Demourant, Poussot-Vassal \&
  Sipp]{leclercq2019linear}
{\sc \au{Leclercq, C.}, \au{Demourant, F.}, \au{Poussot-Vassal, C.} \&
  \au{Sipp, D.}} \yr{2019}  \at{Linear iterative method for closed-loop control
  of quasiperiodic flows}.  \jt{J. Fluid Mech.}  \bvol{868},  \pg{26--65}.

\bibitem[Liu {\em et~al.\/}(2016)Liu, G{\'o}mez \& Theofilis]{liu2016linear}
{\sc \au{Liu, Q.}, \au{G{\'o}mez, F.} \& \au{Theofilis, V.}} \yr{2016}
  \at{Linear instability analysis of low-${R}e$ incompressible flow over a long
  rectangular finite-span open cavity}.  \jt{J. Fluid Mech.}  \bvol{799}.

\bibitem[Lusk {\em et~al.\/}(2012)Lusk, Cattafesta \& Ukeiley]{lusk2012leading}
{\sc \au{Lusk, T.}, \au{Cattafesta, L.~N.} \& \au{Ukeiley, L.~S.}} \yr{2012}
  \at{Leading edge slot blowing on an open cavity in supersonic flow}.
  \jt{Exp. Fluids}  \bvol{53}~(1),  \pg{187--199}.

\bibitem[Marques~Ribeiro {\em et~al.\/}(2020)Marques~Ribeiro, Yeh \&
  Taira]{Ribeiro:PRF20}
{\sc \au{Marques~Ribeiro, J.~H.}, \au{Yeh, C.-A.} \& \au{Taira, K.}} \yr{2020}
  \at{Randomized resolvent analysis (accepted)}.  \jt{Phys. Rev. Fluids} .

\bibitem[Maull \& East(1963)]{maull1963three}
{\sc \au{Maull, D.~J.} \& \au{East, L.~F.}} \yr{1963}  \at{Three-dimensional
  flow in cavities}.  \jt{J. Fluid Mech.}  \bvol{16}~(4),  \pg{620--632}.

\bibitem[McGrath \& Shaw(1996)]{mcgrath1996active}
{\sc \au{McGrath, S.~F.} \& \au{Shaw, L.~L.}} \yr{1996}  \bt{Active control of
  shallow cavity acoustic resonance}. {AIAA} Paper 96-1949.

\bibitem[Mcgregor \& White(1970)]{mcgregor1970drag}
{\sc \au{Mcgregor, O.~W.} \& \au{White, R.~A.}} \yr{1970}  \at{Drag of
  rectangular cavities in supersonic and transonic flow including the effects
  of cavity resonance}.  \jt{AIAA J.}  \bvol{8}~(11),  \pg{1959--1964}.

\bibitem[McKeon \& Sharma(2010)]{mckeon2010critical}
{\sc \au{McKeon, B.~J.} \& \au{Sharma, A.~S.}} \yr{2010}  \at{A critical-layer
  framework for turbulent pipe flow}.  \jt{J. Fluid Mech.}  \bvol{658},
  \pg{336--382}.

\bibitem[Murray {\em et~al.\/}(2009)Murray, S{\"a}llstr{\"o}m \&
  Ukeiley]{murray2009properties}
{\sc \au{Murray, N.}, \au{S{\"a}llstr{\"o}m, E.} \& \au{Ukeiley, L.~S.}}
  \yr{2009}  \at{Properties of subsonic open cavity flow fields}.  \jt{Phys.
  Fluids}  \bvol{21}~(9),  \pg{095103}.

\bibitem[Nakashima {\em et~al.\/}(2017)Nakashima, Fukagata \&
  Luhar]{nakashima2017assessment}
{\sc \au{Nakashima, S.}, \au{Fukagata, K.} \& \au{Luhar, M.}} \yr{2017}
  \at{Assessment of suboptimal control for turbulent skin friction reduction
  via resolvent analysis}.  \jt{J. Fluid Mech.}  \bvol{828},  \pg{496--526}.

\bibitem[Picella {\em et~al.\/}(2018)Picella, Loiseau, Lusseyran, Robinet,
  Cherubini \& Pastur]{picella2018successive}
{\sc \au{Picella, F.}, \au{Loiseau, J-Ch.}, \au{Lusseyran, F.}, \au{Robinet,
  J-Ch.}, \au{Cherubini, S.} \& \au{Pastur, L.}} \yr{2018}  \at{Successive
  bifurcations in a fully three-dimensional open cavity flow}.  \jt{J. Fluid
  Mech.}  \bvol{844},  \pg{855--877}.

\bibitem[Plumblee {\em et~al.\/}(1962)Plumblee, Gibson \&
  Lassiter]{plumblee1962theoretical}
{\sc \au{Plumblee, H.~E.}, \au{Gibson, J.~S.} \& \au{Lassiter, L.~W.}}
  \yr{1962}  \bt{A theoretical and experimental investigation of the acoustic
  response of cavities in an aerodynamic flow}. {\em Tech. Rep.\/}.
  \org{Lockheed Aircraft Corp Marietta GA}.

\bibitem[Rizzetta \& Visbal(2003)]{rizzetta2003large}
{\sc \au{Rizzetta, D.~P.} \& \au{Visbal, M.~R.}} \yr{2003}  \at{Large-eddy
  simulation of supersonic cavity flowfields including flow control}.  \jt{AIAA
  J.}  \bvol{41}~(8),  \pg{1452--1462}.

\bibitem[Rockwell \& Naudascher(1979)]{rockwell1979self}
{\sc \au{Rockwell, D.} \& \au{Naudascher, E.}} \yr{1979}  \at{Self-sustained
  oscillations of impinging free shear layers}.  \jt{Annu. Rev. Fluid Mech.}
  \bvol{11}~(1),  \pg{67--94}.

\bibitem[Rossiter(1964)]{Rossiter1964}
{\sc \au{Rossiter, J.~E.}} \yr{1964}  \bt{Wind-tunnel experiments on the flow
  over rectangular cavities at subsonic and transonic speeds}. {\em Tech.
  Rep.\/} 3438.  \org{Aeronautical Research Council Reports and Memoranda}.

\bibitem[Rowley {\em et~al.\/}(2009)Rowley, Mezi{\'c}, Bagheri, Schlatter \&
  Henningson]{rowley2009spectral}
{\sc \au{Rowley, C.~W.}, \au{Mezi{\'c}, I.}, \au{Bagheri, S.}, \au{Schlatter,
  P.} \& \au{Henningson, D.~S.}} \yr{2009}  \at{Spectral analysis of nonlinear
  flows}.  \jt{J. Fluid Mech.}  \bvol{641},  \pg{115--127}.

\bibitem[Rowley \& Williams(2006)]{rowley2006dynamics}
{\sc \au{Rowley, C.~W.} \& \au{Williams, D.~R.}} \yr{2006}  \at{Dynamics and
  control of high {R}eynolds number flow over open cavities}.  \jt{Annu. Rev.
  Fluid Mech.}  \bvol{38},  \pg{251--276}.

\bibitem[Sarno \& Franke(1994)]{sarno1994suppression}
{\sc \au{Sarno, R.~L.} \& \au{Franke, M.~E.}} \yr{1994}  \at{Suppression of
  flow-induced pressure oscillations in cavities}.  \jt{J. Aircr.}
  \bvol{31}~(1),  \pg{90--96}.

\bibitem[Schmid(2010)]{schmid2010dynamic}
{\sc \au{Schmid, P.~J.}} \yr{2010}  \at{Dynamic mode decomposition of numerical
  and experimental data}.  \jt{J. Fluid Mech.}  \bvol{656},  \pg{5--28}.

\bibitem[Schmid \& Henningson(2012)]{schmid2012stability}
{\sc \au{Schmid, P.~J.} \& \au{Henningson, D.~S.}} \yr{2012} {\em Stability and
  transition in shear flows\/}.  \publ{Springer}.

\bibitem[Schmidt {\em et~al.\/}(2017)Schmidt, Towne, Colonius, Cavalieri,
  Jordan \& Br{\`e}s]{Schmidt:JFM2017}
{\sc \au{Schmidt, O.~T.}, \au{Towne, A.}, \au{Colonius, T.}, \au{Cavalieri, A.
  V.~G.}, \au{Jordan, P.} \& \au{Br{\`e}s, G.~A.}} \yr{2017}  \at{Wavepackets
  and trapped acoustic modes in a turbulent jet: coherent structure eduction
  and global stability}.  \jt{J. Fluid Mech.}  \bvol{825},  \pg{1153--1181}.

\bibitem[Shaw(1998)]{shaw1998active}
{\sc \au{Shaw, L.}} \yr{1998}  \bt{Active control for cavity acoustics}. {AIAA}
  Paper 1998-2347.

\bibitem[Sun {\em et~al.\/}(2019{\natexlab{{\em a\/}}})Sun, Liu, Cattafesta,
  Ukeiley \& Taira]{sun2018effects}
{\sc \au{Sun, Y.}, \au{Liu, Q.}, \au{Cattafesta, L.~N.}, \au{Ukeiley, L.~S.} \&
  \au{Taira, K.}} \yr{2019{\natexlab{{\em a\/}}}}  \at{Effects of sidewalls and
  leading-edge blowing on flows over long rectangular cavities}.  \jt{AIAA J.}
  \bvol{57}~(1),  \pg{106--119}.

\bibitem[Sun {\em et~al.\/}(2019{\natexlab{{\em b\/}}})Sun, Liu, N., Ukeiley \&
  Taira]{sun2019resolvent}
{\sc \au{Sun, Y.}, \au{Liu, Q.}, \au{N., Cattafesta~L.}, \au{Ukeiley, L.~S.} \&
  \au{Taira, K.}} \yr{2019{\natexlab{{\em b\/}}}}  \at{Resolvent analysis of
  compressible laminar and turbulent cavity flows}.  \jt{AIAA J. (accepted)} .

\bibitem[Sun {\em et~al.\/}(2017)Sun, Taira, Cattafesta \&
  Ukeiley]{sun2017biglobal}
{\sc \au{Sun, Y.}, \au{Taira, K.}, \au{Cattafesta, L.~N.} \& \au{Ukeiley,
  L.~S.}} \yr{2017}  \at{Biglobal instabilities of compressible open-cavity
  flows}.  \jt{J. Fluid Mech.}  \bvol{826},  \pg{270--301}.

\bibitem[Taira {\em et~al.\/}(2017)Taira, Brunton, Dawson, Rowley, Colonius,
  McKeon, Schmidt, Gordeyev, Theofilis \& Ukeiley]{Taira:AIAAJ17}
{\sc \au{Taira, K.}, \au{Brunton, S.~L.}, \au{Dawson, S.}, \au{Rowley, C.~W.},
  \au{Colonius, T.}, \au{McKeon, B.~J.}, \au{Schmidt, O.~T.}, \au{Gordeyev,
  S.}, \au{Theofilis, V.} \& \au{Ukeiley, L.~S.}} \yr{2017}  \at{Modal analysis
  of fluid flows: An overview}.  \jt{AIAA J.}  \bvol{55}~(12),
  \pg{4013--4041}.

\bibitem[Taira {\em et~al.\/}(2020)Taira, Hemati, Brunton, Sun, Duraisamy,
  Bagheri, Dawson \& Yeh]{Taira:AIAAJ20}
{\sc \au{Taira, K.}, \au{Hemati, M.~S.}, \au{Brunton, S.~L.}, \au{Sun, Y.},
  \au{Duraisamy, K.}, \au{Bagheri, S.}, \au{Dawson, S. T.~M.} \& \au{Yeh,
  C.-A.}} \yr{2020}  \at{Modal analysis of fluid flows: applications and
  outlook}.  \jt{AIAA J.} .

\bibitem[Theofilis(2011)]{theofilis2011global}
{\sc \au{Theofilis, V.}} \yr{2011}  \at{Global linear instability}.  \jt{Annu.
  Rev. Fluid Mech.}  \bvol{43},  \pg{319--352}.

\bibitem[Toro {\em et~al.\/}(1994)Toro, Spruce \& Speares]{Toro:94}
{\sc \au{Toro, E.~F.}, \au{Spruce, M.} \& \au{Speares, W.}} \yr{1994}
  \at{Restoration of the contact surface in the {HLL}-{R}iemann solver}.
  \jt{Shock Waves}  \bvol{4},  \pg{25--34}.

\bibitem[Trefethen \& Embree(2005)]{trefethen2005spectra}
{\sc \au{Trefethen, L.~N.} \& \au{Embree, M.}} \yr{2005} {\em Spectra and
  pseudospectra: the behavior of nonnormal matrices and operators\/}.
  \publ{Princeton University Press}.

\bibitem[Trefethen {\em et~al.\/}(1993)Trefethen, Trefethen, Reddy \&
  Driscoll]{trefethen1993hydrodynamic}
{\sc \au{Trefethen, L.~N.}, \au{Trefethen, A.~E.}, \au{Reddy, S.~C.} \&
  \au{Driscoll, T.~A.}} \yr{1993}  \at{Hydrodynamic stability without
  eigenvalues}.  \jt{Science}  \bvol{261}~(5121),  \pg{578--584}.

\bibitem[Ukeiley {\em et~al.\/}(2004)Ukeiley, Ponton, Seiner \&
  Jansen]{ukeiley2004suppression}
{\sc \au{Ukeiley, L.~S.}, \au{Ponton, M.~K.}, \au{Seiner, J.~M.} \& \au{Jansen,
  B.}} \yr{2004}  \at{Suppression of pressure loads in cavity flows}.  \jt{AIAA
  J.}  \bvol{42}~(1),  \pg{70--79}.

\bibitem[Vakili \& Gauthier(1994)]{vakili1994control}
{\sc \au{Vakili, A.~D.} \& \au{Gauthier, C.}} \yr{1994}  \at{Control of cavity
  flow by upstream mass-injection}.  \jt{J. Aircr.}  \bvol{31}~(1),
  \pg{169--174}.

\bibitem[Vreman(2004)]{Vreman:PF04}
{\sc \au{Vreman, A.~W.}} \yr{2004}  \at{An eddy-viscosity subgrid-scale model
  for turbulent shear flow: algebraic theory and applications}.  \jt{Phys.
  Fluids}  \bvol{16}~(10),  \pg{3670--3681}.

\bibitem[Williams {\em et~al.\/}(2007)Williams, Cornelius \&
  Rowley]{williams2007supersonic}
{\sc \au{Williams, D.~R.}, \au{Cornelius, D.} \& \au{Rowley, C.~W.}} \yr{2007}
  \at{Supersonic cavity response to open-loop forcing}.  \bt{In {\em Active
  Flow Control\/}},  \pg{pp. 230--243}.  \publ{Springer}.

\bibitem[Yeh {\em et~al.\/}(2020)Yeh, Benton, Taira \& Garmann]{Yeh:PRF20}
{\sc \au{Yeh, C.-A.}, \au{Benton, S.~I.}, \au{Taira, K.} \& \au{Garmann,
  D.~J.}} \yr{2020} Resolvent analysis of an airfoil laminar separation bubble
  at ${R}e = 500,000$. In review.

\bibitem[Yeh \& Taira(2019)]{yeh2018resolvent}
{\sc \au{Yeh, C.-A.} \& \au{Taira, K.}} \yr{2019}  \at{Resolvent-analysis-based
  design of airfoil separation control}.  \jt{J. Fluid Mech.}  \bvol{867},
  \pg{572--610}.

\bibitem[Zhang {\em et~al.\/}(2019)Zhang, Sun, Arora, Cattafesta, Taira \&
  Ukeiley]{zhang2019suppression}
{\sc \au{Zhang, Y.}, \au{Sun, Y.}, \au{Arora, N.}, \au{Cattafesta, L.~N.},
  \au{Taira, K.} \& \au{Ukeiley, L.~S.}} \yr{2019}  \at{Suppression of cavity
  flow oscillations via three-dimensional steady blowing}.  \jt{AIAA J.}
  \bvol{57}~(1),  \pg{90--105}.

\end{thebibliography}
\end{document}